\documentclass{article} 
\usepackage{iclr2027_conference,times}

\usepackage{amsmath,amsfonts,bm}

\def\eqref#1{equation~\ref{#1}}

\def\1{\bm{1}}

\DeclareMathAlphabet{\mathsfit}{\encodingdefault}{\sfdefault}{m}{sl}
\SetMathAlphabet{\mathsfit}{bold}{\encodingdefault}{\sfdefault}{bx}{n}

\usepackage{hyperref}
\usepackage{url}
\usepackage{booktabs}
\usepackage{tabularx}
\usepackage{multirow}
\usepackage{graphicx}
\usepackage{placeins}
\usepackage{fvextra}
\fvset{fontfamily=cmtt}
\usepackage{xurl}
\usepackage{needspace}
\usepackage{microtype}

\newcommand{\mainfloatlayout}{%
  \setlength{\textfloatsep}{12pt plus 2pt minus 2pt}%
  \setlength{\floatsep}{10pt plus 2pt minus 2pt}%
  \setlength{\intextsep}{12pt plus 2pt minus 2pt}%
  \renewcommand{\floatpagefraction}{0.85}%
  \setcounter{topnumber}{3}%
  \setcounter{totalnumber}{4}%
}
\newcommand{\appendixfloatlayout}{%
  \setlength{\textfloatsep}{2\baselineskip}%
  \setlength{\floatsep}{2\baselineskip}%
  \setlength{\intextsep}{2\baselineskip}%
  \setlength{\belowcaptionskip}{6pt}%
  \renewcommand{\floatpagefraction}{0.5}%
  \setcounter{topnumber}{2}%
  \setcounter{totalnumber}{3}%
}
\mainfloatlayout

\title{Take It from Here: Evaluating\\Coding Agents on Project Takeover}

\author{%
\normalfont
Dingxian Wang\textsuperscript{1,}\thanks{These authors contributed equally to this work.}\quad
Zhe Fu\textsuperscript{2,}\footnotemark[1]\quad
Jiaqi Deng\textsuperscript{2,}\footnotemark[1]\\
Chao Yin\textsuperscript{1}\quad Ganesh Prabhu\textsuperscript{1}\quad Karthik Gomadam Rajagopal\textsuperscript{1}\\
Darwin Yi\textsuperscript{3}\quad Huan Huo\textsuperscript{2,}\thanks{Corresponding author: Huan Huo.}\quad Guandong Xu\textsuperscript{2,4}\quad Xianzhi Wang\textsuperscript{2}\\
Andrew Rabinovich\textsuperscript{1}\quad Zhao Chen\textsuperscript{1}\\[0.5em]
{\small \textsuperscript{1}Upwork Inc\quad \textsuperscript{2}University of Technology Sydney}\\
{\small \textsuperscript{3}Scale AI\quad \textsuperscript{4}The Education University of Hong Kong}
}
\hypersetup{pdfauthor={Dingxian Wang, Zhe Fu, Jiaqi Deng, Chao Yin, Ganesh Prabhu, Karthik Gomadam Rajagopal, Darwin Yi, Huan Huo, Guandong Xu, Xianzhi Wang, Andrew Rabinovich, Zhao Chen}}

\iclrfinalcopy 
\begin{document}
\microtypesetup{activate=false}
\raggedbottom

\maketitle
\lhead{Preprint}

\begin{abstract}

Coding agents increasingly collaborate on long-term projects. When taking over a project, they need to recover currently applicable requirements from the interaction history and update them in response to new requests. However, most coding benchmarks do not include long interaction histories, and some evaluations that consider history still do not cover the full process from requirement reconstruction to project delivery. We therefore provide agents with the interaction history, current request, and pre-request project environment, ask them to make the requested changes, and assess takeover performance through delivery tests. Across six model--harness configurations and 37 targets eligible for executable validation, the mean complete-delivery rate under Full History is only 4.95\%. We then use requirement atoms to independently diagnose a larger set of takeover targets, examining relevant information identification, pre-request state reconstruction, and current request processing. We further supply correct pre-request or target states to examine the remaining difficulties in requirement processing and project modification. The diagnostics show that none of the six configurations produces fully correct post-request requirement states, and a case illustrates omission of previously recovered constraints during updating. With correct target states supplied, the mean complete-delivery rate is 8.11\%, with build and regression pass rates of 96.40\% and 90.99\%, but target-test success remains 9.01\%. These results show difficulties in both forming target requirement states and implementing supplied targets in existing projects. Based on these findings, we introduce ReqMemBench, which combines requirement diagnostics with delivery validation to support systematic evaluation of project takeover.
\end{abstract}

\section{Introduction}
\label{sec:introduction}

Coding agents have progressed from generating standalone code snippets \citep{chen2021codex} to browsing files, modifying implementations, and running tests in existing codebases \citep{yang2024sweagent,wang2025openhands}. When serving as assistants on long-term projects, they need to understand new user requests and build on existing work to move the project forward \citep{le2026sweevo,orlanski2026slopcode}. However, the information needed to interpret the current request is often scattered across earlier interactions \citep{wu2025longmemeval,laban2025llms}. Therefore, Coding agents must extract task-relevant requirements from long interaction histories, determine how the new request changes those requirements, and then produce the required deliverables.

Existing coding benchmarks mainly assess code generation and repository-level tasks, typically without long interaction histories \citep{jimenez2024swebench,miserendino2025swelancer}. Although some studies examine history use, requirement updates, or intermediate steps, they do not cover the full process at a specified takeover point: reconstructing the current requirement state from history and delivering the requested work \citep{wu2025longmemeval,hu2025memoryagentbench,liu2026locoeval,zhou2026swerpg}. We evaluate this process across different deliverable types.

\begin{figure}[!t]
\centering
\includegraphics[width=\linewidth]{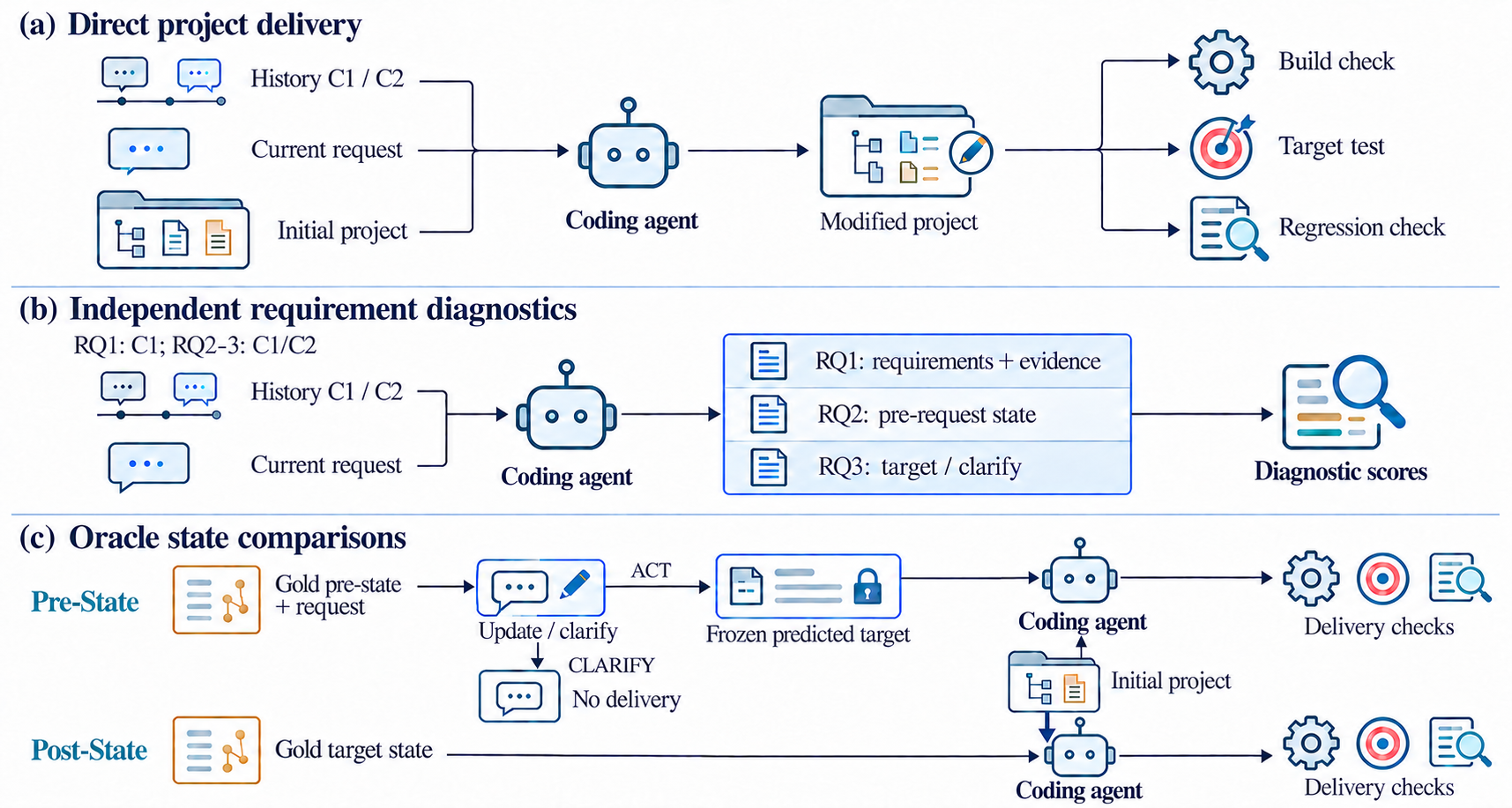}
\caption{Project takeover experiments. (a) Direct delivery from history, the current request, and the initial project, followed by three delivery checks. (b) Independent requirement diagnostics (RQ1: C1 only). (c) Oracle Pre-State predicts the target from the Gold pre-state, whereas Oracle Post-State supplies the Gold target directly, followed by delivery evaluation. Diagnostic responses are not inputs to the main experiment.}
\label{fig:overview}
\end{figure}

We construct takeover tasks from 44 projects' long-term interactions and test direct delivery on 37 targets from 7 of them (Figure~\ref{fig:overview}). Agents receive the history, current request, and pre-request project environment and submit modified projects for testing. We compare Full History (C1) with shorter, filtered Relevant History (C2) to examine whether shortening context improves delivery. Across six model--harness configurations, mean complete-delivery rates are only 4.95\% and 2.70\%, respectively.

To investigate takeover difficulties, we link multi-turn descriptions through requirement atoms: semantically complete requirements that can evolve independently. On 349 targets, three diagnostics (RQ1--RQ3) assess relevant information identification, pre-request state reconstruction, and current request updating. For requests with sufficient information to proceed, no configuration produces a fully correct target state under either history condition. A case further shows that updating can omit previously recovered constraints that still apply, highlighting the difficulty of forming complete and accurate updated states.

We then use Oracle Pre-State and Oracle Post-State to examine whether target requirements become actual project changes. Pre-State supplies the correct pre-request state for the agent to update and implement; Post-State directly supplies the correct target. Under Post-State, build and regression pass rates are 96.40\% and 90.99\%, respectively, but target-test success is only 9.01\%. Thus, correct targets still do not ensure implementation of the requested changes.

These findings motivate ReqMemBench, which links requirement diagnostics with project delivery in two phases. Agents first produce and freeze requirement responses from the history and current request. For execution-eligible targets judged actionable, they then modify the initial project using those responses and undergo delivery validation. Our contributions are:

\begin{itemize}
\item We identify difficulties in both requirement understanding and project modification during takeover: agents struggle to form complete and accurate target requirement states from long interaction histories, and directly supplying correct targets still does not ensure implementation of the requested changes.
\item We establish a project-takeover evaluation framework combining direct delivery, independent requirement diagnostics, and Oracle state comparisons to assess historical requirement understanding, target-state formation, and implementation in existing projects.
\item We instantiate this evaluation design as ReqMemBench across multiple deliverable types, with 349 takeover targets from 44 projects and initial project environments and delivery validators for 37 targets, supporting joint evaluation of requirement understanding and project delivery.
\end{itemize}

\section{Related Work}
\label{sec:related-work}

Existing coding benchmarks assess both function generation from specifications \citep{chen2021codex,austin2021mbpp,zhuo2025bigcodebench} and code completion using information across multiple files \citep{ding2023crosscodeeval,liu2024repobench}. At the project level, SWE-bench and SWE-Lancer examine agents' ability to solve real engineering tasks \citep{jimenez2024swebench,miserendino2025swelancer}, while SWE-EVO and ProjDevBench further examine continuous evolution of existing software and complete project development \citep{le2026sweevo,lu2026projdevbench}. These evaluations typically start from a given task description and code environment. In project takeover, however, the requirements governing the current task are scattered across earlier interactions, so agents must first determine which historical information still applies.

Long-term memory benchmarks also examine recall and reasoning over conversational histories \citep{maharana2024locomo}. LongMemEval tests temporal reasoning, knowledge updating, and abstention when information is insufficient through multi-session question answering \citep{wu2025longmemeval}. MemoryAgentBench incrementally supplies information to assess retrieval, learning, understanding, and discarding superseded facts \citep{hu2025memoryagentbench}. LongMemEval-V2 examines inference of environmental changes and workflows from past web interactions \citep{wu2026longmemevalv2}. These studies evaluate historical retrieval and updating, whereas takeover also requires integrating that information into requirements for the current delivery.

In coding, interactive benchmarks assess how agents revise code through multi-turn feedback or stepwise requirement refinement \citep{han2025convcodeworld,miao2026recodeh,zhan2025sreval}. Sequential-task benchmarks examine reuse of prior experience and code quality under evolving requirements \citep{joshi2025swebenchcl,zhu2026swecontextbench,orlanski2026slopcode}. LoCoEval introduces conflicts and updates in long conversations, then separately evaluates topic identification, information extraction, and function generation after the complete conversation \citep{liu2026locoeval}. SWE-RPG diagnoses requirement and planning errors against reference answers, while RigorBench evaluates planning, verification, and recovery in execution traces \citep{zhou2026swerpg,madiraju2026rigorbench}. These studies cover parts of the process. We evaluate project takeover at a specified point: reconstructing and updating applicable requirements from long histories, then delivering the requested changes. Combining requirement-state diagnostics and delivery validation also lets us examine whether agents implement explicit target requirements in an ongoing project.

\section{Methods and Data Construction}
\label{sec:methods}
\label{sec:formulation}

We use requests made midway through a project as evaluation units. This section defines the takeover task and delivery evaluation, introduces the representation and scoring for independent requirement diagnostics, and describes data construction.

\subsection{Project Takeover Task and Delivery Evaluation}
\label{sec:delivery-task}

In direct delivery, agents receive the prior history, current request, and pre-request project environment, then submit the modified project (Figure~\ref{fig:overview}(a); history conditions in Section~\ref{sec:main-experiment}). Three checks evaluate the result. The build check assesses code buildability or basic project-file validity. The target test checks implementation of the requested change, while the regression check assesses preservation of specified historical constraints that remain valid. Complete delivery requires all three checks to pass:

\begin{equation}
\mathrm{DeliveryRate}
=\frac{\sum_{i\in\mathcal T_{\mathrm{exec}}}
\mathbf 1[B_i\land T_i\land R_i]}
{|\mathcal T_{\mathrm{exec}}|},
\label{eq:delivery-rate}
\end{equation}

Here, $\mathcal T_{\mathrm{exec}}$ is the predefined executable target set, and $B_i,T_i,R_i$ indicate the three check outcomes. Incomplete runs and missing deliverables count as failures. Infrastructure or validator errors remain pending until resolved, after which rates use the full target set. We also report individual check pass rates to identify unmet delivery requirements. Decision rules and error handling are in Appendix \ref{app:c2}.

\subsection{Requirement Understanding Diagnostics}
\label{sec:requirement-diagnostics}

Requirement understanding affects agents' delivery performance \citep{mu2024clarifygpt,zhou2026swerpg}, yet delivery checks do not directly reveal omissions or errors in that understanding. We therefore conduct independent requirement diagnostics (Figure~\ref{fig:overview}(b)). Because one message may concern several requirements and a requirement may change across interactions, we link these descriptions through requirement atoms: semantically complete requirements that can evolve independently. This follows the requirements engineering practice of tracing requirement changes and their sources \citep{nuseibeh2000requirements,gotel1994analysis}.

Each requirement atom has five state components. Attributes record its settings, while lifecycle status and scope indicate whether it still applies. Ambiguity records unresolved questions, and implementation evidence records earlier implementation claims or validation results. For requirements involved in the current request, $G_q^-$ denotes their pre-request state and $G_q^+$ the target state, which incorporates current changes while retaining valid earlier content. New requirements appear only in $G_q^+$; neither state uses subsequent implementation or runtime feedback. Definitions, fields, and replay rules are in Appendix \ref{app:a1}.

Three progressive diagnostics assess relevant requirement and evidence identification (RQ1), pre-request state reconstruction (RQ2), and current request updating or clarification (RQ3). In RQ3, when information is sufficient to determine target requirements, the agent selects ACT and outputs a complete post-request state. If critical information is missing, it selects CLARIFY, identifies the missing information, and asks clarification questions. These decisions apply to diagnostics, while the main experiment uses direct delivery (Section~\ref{sec:delivery-task}). Table~\ref{tab:diagnostic-questions} lists the evaluation objects and metrics; Appendix \ref{app:a2} gives a state-update example.

\begin{table}[htbp]
\centering\small
\setlength{\tabcolsep}{3pt}
\renewcommand{\arraystretch}{1.08}
\caption{What the requirement diagnostics evaluate.}
\label{tab:diagnostic-questions}
\begin{tabularx}{\linewidth}{@{}>{\raggedright\arraybackslash}p{0.22\linewidth}>{\raggedright\arraybackslash}p{0.32\linewidth}>{\raggedright\arraybackslash}X@{}}
\toprule
Diagnostic & \shortstack[c]{What is\\evaluated} & Core metrics \\
\midrule
RQ1: Relevant information identification & Historical requirements and evidence relevant to the current task & Precision, Recall, and F1 computed separately for requirements and evidence \\
RQ2: Pre-request state reconstruction & Pre-request states of relevant requirements, $G_q^-$ & Attribute Reconstruction Score for matched requirements; Requirement Coverage \\
RQ3: Request update or clarification & ACT/CLARIFY decision, and $G_q^+$ or material missing information and clarification questions & Decision Accuracy, Exact Post-State Accuracy, Post-State Score, Blocking-Issue F1, and Question Validity \\
\bottomrule
\end{tabularx}
\end{table}

Predictions may express the same meaning as the gold-standard answer (Gold) in different words. We therefore use a semantic judge to determine their relationships, after which the program performs one-to-one requirement matching and compares field values. RQ1 and RQ2 match requirements independently: RQ1 evaluates requirement identification and scores evidence selection independently of requirement matching, while RQ2 reports attribute quality on matched requirements and Requirement Coverage to distinguish reconstruction errors from omissions. The program detects conflicting relation labels across diagnostics, updates matching after human review, and recomputes affected metrics (Appendix \ref{app:c3}). Matching rules and judge settings are given in Appendix \ref{app:c1} and Appendix \ref{app:d1}, respectively.

\noindent\begin{minipage}{\linewidth}
For requests that should be acted upon, let $\widehat{\mathcal R}_q$ and $\mathcal R_q$ denote the predicted and Gold requirement sets, respectively, and $M_q^+$ their one-to-one matching. For each matched requirement $r$, $s_r\in[0,1]$ is its mean score over applicable state dimensions. The Post-State Score is

\begin{equation}
\mathrm{PostState}_q=
\frac{2\sum_{r\in M_q^+}s_r}
{|\widehat{\mathcal R}_q|+|\mathcal R_q|}.
\label{eq:post-state-score}
\end{equation}
\end{minipage}

This metric gives partial credit for partially correct states and penalizes missing or extra requirements. We also report Exact Post-State Accuracy, which requires a correct ACT decision, no missing or extra requirements, correct values for all applicable state fields with no extra state dimensions, and a list of attribute names to remove that matches Gold.

For requests that require clarification, Blocking-Issue F1 measures the identification of critical missing information. Question Validity is evaluated only for blocking issues matched to Gold and measures whether the proposed questions can obtain the needed information. An incorrect action decision gives the Gold branch a content score of 0. Field scoring and null-value rules are given in Appendix \ref{app:c1}, and aggregation rules in Appendix \ref{app:c2}.

\subsection{Data Construction and Validation}
\label{sec:construction}

The materials cover 44 ongoing collaborative projects involving software, documents, and PCB design. Counting each normalized message as one turn, complete project histories range from 18 to 1,075 turns (mean 181.4; median 110).

We annotate requirement atoms, change events, and source messages, replay events chronologically to construct each atom's state graph, and select history-dependent requests with evaluable requirement changes, yielding 349 diagnostic targets. Each target includes the current request, preceding history, and task-relevant Gold pre-request and post-request states or clarification annotations. Histories end before the current request, and Gold excludes subsequent implementation and runtime feedback.

Diagnostic inputs average 189.8 turns under C1 (median 118) and 9.7 under C2 (median 6). The construction process and complete length distributions are provided in Appendix \ref{app:b2} and Appendix \ref{app:b1}, respectively.

Programs check message references, event order, state transitions, and history boundaries, and an agent reviews RQ2/RQ3 Gold structure and semantics. Four human reviewers in two pairs cover all 44 projects, with two independent reviews per project. Agreements are accepted; disagreements are reviewed again. Gold review and semantic-score verification are described in Appendix \ref{app:b3} and Appendix \ref{app:c3}, respectively.

Delivery targets additionally require reproducible pre-request environments and executable acceptance tests. Initial environments must pass build and regression checks but fail the target test. These criteria yield 37 delivery targets from 7 projects; selection and validation scope are detailed in Appendix \ref{app:b3}.

\section{Experimental Design}
\label{sec:experimental-design}
\label{sec:experiments}

The three experiments use the same six model--harness configurations to evaluate direct delivery, requirement understanding, and project modification with correct-state inputs.

The six configurations comprise GPT-6 Astra, GPT-6 Sol, GPT-5.6 Sol, and GPT-5.6 Terra with Codex CLI, and Claude Opus 5.5 and Claude Sonnet 5 with Claude Code, all using high reasoning effort. Each configuration, target, and condition has one independent run in a new process, and delivery experiments also use separate project environments. Requirement diagnostics have a 900-second time limit, and direct delivery has a 1,800-second limit. Specific versions, tool permissions, prompts, and semantic judge settings are given in Appendix \ref{app:d1} and Appendix \ref{app:d3}.

\subsection{Project Takeover Experiment}
\label{sec:main-experiment}

The main experiment follows the direct-delivery procedure in Section~\ref{sec:delivery-task}, using Full History (C1) as the main condition and Relevant History (C2) as the comparison. C1 contains all eligible messages before the takeover point. Using requirement annotations, C2 retains the historical descriptions, earlier revisions, and supporting messages for requirements involved in the current request, removes unrelated messages, and preserves the original order. Its construction is described in Appendix \ref{app:b2}. Across the 37 delivery targets, the mean history lengths are 186.9 turns under C1 and 8.7 under C2.

Both conditions share targets, current requests, initial environments, tool permissions, budgets, and validators within each configuration; only visible history changes. Motivated by difficulties reported in long-context evaluations \citep{bai2024longbench,hsieh2024ruler}, C2 examines whether shortening context improves delivery. Since it also changes information composition, the comparison measures the filter's overall effect. Rates use all 37 targets (Section~\ref{sec:construction}) per configuration and condition, and cross-configuration results average the six configuration-level rates.

\subsection{Requirement Understanding Diagnostic Experiment}
\label{sec:diagnostics-experiment}

The requirement diagnostics use 349 targets from 44 projects, providing only the history, current request, and response specification, without project environment access. The model submits a single structured response following Section~\ref{sec:requirement-diagnostics}. This response is frozen and scored independently and is not supplied as an additional input to the main experiment.

RQ1 uses C1 only, while RQ2 and RQ3 compare C1 and C2. For every configuration and condition, RQ3 decision metrics use all 349 targets, Exact Post-State Accuracy and Post-State Score both use all 312 Gold-ACT targets, and clarification metrics use the 37 Gold-CLARIFY targets. To distinguish action-selection errors from difficulties in constructing the target state, we also report Post-State Score conditional on a correct ACT decision.

Because the two classes are imbalanced, decision evaluation reports accuracy, recall for each class, balanced accuracy, and an always-ACT reference. Content metrics are computed per target and then averaged over their applicable sets. Results are aggregated separately for each configuration before taking an equal-weight average across the six configurations. Matching, metric applicability, and failure scoring rules are given in Appendices \ref{app:c1}–\ref{app:c2}.

\subsection{Oracle State Comparison Experiment}
\label{sec:oracle-experiment}

Oracle comparisons examine project modification from correct-state inputs on the same six configurations and 37 targets (Figure~\ref{fig:overview}(c)), with initial environments, tool permissions, coding budgets, and validators held fixed within each configuration.

Oracle Pre-State supplies $G_q^-$ and the current request, bypassing historical reconstruction. The model updates requirements or requests clarification. For ACT, its predicted target state is frozen and passed with the initial environment to coding for project modification and submission; CLARIFY does not enter coding. Oracle Post-State directly supplies Gold $G_q^+$ and the initial environment for delivery. In both conditions, the requirement input to coding contains only references, summaries, and final state fields, from predicted states in Pre-State and Gold states in Post-State (Appendix \ref{app:d2}).

Within each configuration, we compare Post-State with direct C1 delivery and examine both Oracle conditions through the checks in Section~\ref{sec:delivery-task}. Post-State also changes the coding input relative to C1, while Pre-State and Post-State differ in state sources and whether an action decision precedes coding. Rates for complete delivery and each check use all 37 targets; non-coding runs and missing submissions count as failures.

\section{Results and Analysis}
\label{sec:results}

\subsection{Can Agents Take Over Ongoing Projects?}
\label{sec:results-delivery}

Table~\ref{tab:delivery-results} shows low complete-delivery rates under both history conditions. Across the six configurations, the mean rate is 4.95\% under Full History (C1) and 2.70\% under Relevant History (C2). On these 37 targets, the current filtering scheme yields a lower observed mean complete-delivery rate than Full History. We next use independent requirement diagnostics to investigate these difficulties. We examine whether agents can identify relevant requirements and evidence, reconstruct their pre-request states, and determine how to act on the current request.

\begin{table}[htbp]
\centering\small
\setlength{\tabcolsep}{3pt}
\renewcommand{\arraystretch}{1.08}
\caption{Complete-delivery rates (\%) under C1/C2 on the same 37 targets. The mean weights all six configurations equally; aggregation rules are in Appendix \ref{app:c2}.}
\label{tab:delivery-results}
\begin{tabularx}{\linewidth}{@{}>{\raggedright\arraybackslash}X>{\raggedright\arraybackslash}X>{\raggedright\arraybackslash}X>{\raggedright\arraybackslash}X@{}}
\toprule
Harness & \shortstack[c]{Model /\\summary} & \shortstack[c]{Full History\\(C1)} & \shortstack[c]{Relevant\\History (C2)} \\
\midrule
Codex & GPT-6 Astra & 10.81\% & 5.41\% \\
 & GPT-6 Sol & 5.41\% & 2.70\% \\
 & GPT-5.6 Sol & 2.70\% & 5.41\% \\
 & GPT-5.6 Terra & 5.41\% & 0.00\% \\
Claude Code & Opus 5.5 & 5.41\% & 2.70\% \\
 & Sonnet 5 & 0.00\% & 0.00\% \\
Overall & \textbf{Mean across all configurations} & \textbf{4.95\%} & \textbf{2.70\%} \\
\bottomrule
\end{tabularx}
\end{table}

\subsection{Where Do Agents Struggle with Requirement Understanding?}
\label{sec:results-diagnostics}

Table~\ref{tab:diagnostic-selected} shows that selecting relevant historical information and reconstructing pre-request states both remain difficult. Under Full History, GPT-6 Astra has an evidence Recall of 0.9211 but a Precision of only 0.2254. Although it identifies much of the Gold evidence, it also selects many messages outside that set. In RQ2, GPT-6 Sol has a Requirement Coverage of 0.6777 but an Attribute Reconstruction Score of only 0.2671, indicating that identifying a requirement does not ensure accurate recovery of its specific settings. Complete diagnostic results are in Appendix \ref{app:e2}.

\begin{table}[htbp]
\centering\small
\setlength{\tabcolsep}{3pt}
\renewcommand{\arraystretch}{1.08}
\caption{Selected RQ1--RQ2 configuration means under C1 (0--1 scale). Scoring rules are in Appendix \ref{app:c1}, with full results and applicable target counts in Appendix \ref{app:e2}.}
\label{tab:diagnostic-selected}
\begin{tabular}{@{}lrrrr@{}}
\toprule
Configuration & \shortstack[c]{Evidence\\Precision} & \shortstack[c]{Evidence\\Recall} & \shortstack[c]{Requirement\\Coverage} & \shortstack[c]{Attribute\\Reconstruction\\Score} \\
\midrule
GPT-6 Astra & 0.2254 & 0.9211 & 0.6689 & 0.2813 \\
GPT-6 Sol & 0.3395 & 0.8620 & 0.6777 & 0.2671 \\
GPT-5.6 Sol & 0.3162 & 0.8182 & 0.6157 & 0.2451 \\
GPT-5.6 Terra & 0.4254 & 0.6670 & 0.4950 & 0.2424 \\
Claude Opus 5.5 & 0.3189 & 0.8364 & 0.5934 & 0.2811 \\
Claude Sonnet 5 & 0.3060 & 0.7461 & 0.5704 & 0.2548 \\
\bottomrule
\end{tabular}
\end{table}

\begin{figure}[htbp]
\centering
\includegraphics[width=1.0\linewidth]{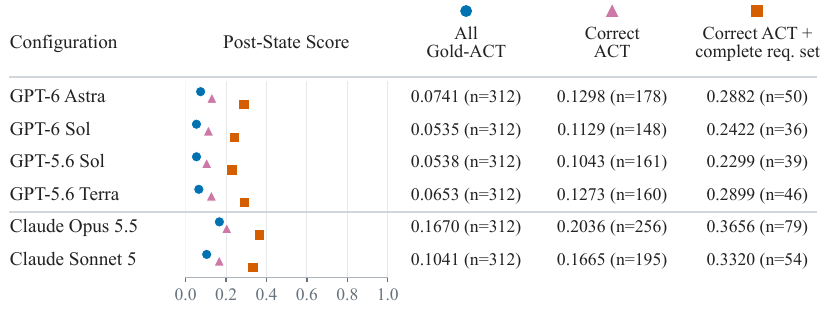}
\caption{Post-request state diagnostics under Full History. Post-State Scores are reported over all Gold-ACT targets, the correct-ACT subset, and the subset with a complete requirement-set match, with target counts shown. The three series describe different evaluation subsets. }
\label{fig:post-state-diagnostics}
\end{figure}

The agents also struggle to form complete and accurate target states after processing the current request. Under both C1 and C2, Exact Post-State Accuracy is 0/312 for every configuration. To examine partially correct predictions, Figure~\ref{fig:post-state-diagnostics} reports Post-State Scores for different target subsets. For example, under Full History, Opus 5.5 scores only 0.2036 on 256 targets where it correctly selects ACT. Even on the 79 targets within this group whose requirement sets fully match Gold, the score is only 0.3656. The other configurations show similar gaps. Thus, a correct action decision and complete requirement identification do not ensure accurate states for those requirements.

Beyond inaccurate state fields, updating can also result in the loss of requirements that the agent has already recovered. In a PCB task, the client asks to move the battery connector to the back of the board, opposite the USB-C port. Opus 5.5 records layout requirements including eight mounting holes and the M2.5 specification in its pre-request state, but omits the entire mechanical-layout requirement from the post-request state in the same response. The current request does not withdraw these requirements, yet they are not retained in the updated target state. Details are in Appendix~\ref{app:e3}.

Together, these results show that forming a complete and accurate target remains difficult: agents can produce incorrect state fields or omit requirements that still apply. We next supply correct pre-request and target states separately to examine the remaining difficulties in request processing and project modification.

\subsection{How Does Providing Correct Requirement States Affect Delivery?}
\label{sec:results-oracle}

With correct target states supplied, Oracle Post-State achieves a mean complete-delivery rate of 8.11\% (Table~\ref{tab:oracle-delivery}), above the 4.95\% under C1 in the main experiment (Table~\ref{tab:delivery-results}), but still below 10\%. To examine which aspects of delivery change, we compare the three checks under main-experiment C1 and Post-State.

The component results in Table~\ref{tab:oracle-delivery} show that, compared with C1 in the main experiment, Post-State increases the build pass rate from 58.56\% to 96.40\% and the regression pass rate from 51.35\% to 90.99\%, whereas target-test success rises only from 6.31\% to 9.01\%. This indicates that, given the correct target, coding agents can identify which requirements should remain unchanged, and that this information also helps them establish a functioning project environment. However, target-test success remains low, indicating that agents still struggle to implement the corresponding changes in code even when given the correct target state.

\begin{table}[htbp]
\caption{Pass rates for individual checks and complete delivery (\%; 37 targets per configuration and condition). C1/C2 denote Full/Relevant History, and Pre/Post denote Oracle Pre-State/Post-State. Means weight their configurations equally; counts and aggregation rules are in Appendix \ref{app:e1}.}
\label{tab:oracle-delivery}
\centering
\begingroup
\fontsize{8}{9.5}\selectfont
\setlength{\tabcolsep}{1pt}
\renewcommand{\arraystretch}{1.1}
\begin{tabular*}{\linewidth}{@{\extracolsep{\fill}}l*{16}{r}@{}}
\toprule
\multirow{2}{*}{Configuration} & \multicolumn{4}{c}{Build} & \multicolumn{4}{c}{Target test} & \multicolumn{4}{c}{Regression} & \multicolumn{4}{c}{Complete delivery} \\
\cmidrule(lr){2-5}\cmidrule(lr){6-9}\cmidrule(lr){10-13}\cmidrule(l){14-17}
 & C1 & C2 & Pre & Post & C1 & C2 & Pre & Post & C1 & C2 & Pre & Post & C1 & C2 & Pre & Post \\
\midrule
GPT-6 Astra & 56.76 & 43.24 & 48.65 & 94.59 & 16.22 & 8.11 & 5.41 & 13.51 & 45.95 & 37.84 & 40.54 & 81.08 & 10.81 & 5.41 & 5.41 & 10.81 \\
GPT-6 Sol & 48.65 & 32.43 & 43.24 & 94.59 & 8.11 & 2.70 & 5.41 & 16.22 & 45.95 & 24.32 & 35.14 & 81.08 & 5.41 & 2.70 & 2.70 & 13.51 \\
GPT-5.6 Sol & 59.46 & 45.95 & 62.16 & 97.30 & 2.70 & 5.41 & 10.81 & 13.51 & 51.35 & 37.84 & 56.76 & 94.59 & 2.70 & 5.41 & 10.81 & 13.51 \\
GPT-5.6 Terra & 54.05 & 37.84 & 51.35 & 91.89 & 5.41 & 0.00 & 2.70 & 5.41 & 48.65 & 27.03 & 45.95 & 89.19 & 5.41 & 0.00 & 2.70 & 5.41 \\
\textbf{Codex mean} & \textbf{54.73} & \textbf{39.86} & \textbf{51.35} & \textbf{94.59} & \textbf{8.11} & \textbf{4.05} & \textbf{6.08} & \textbf{12.16} & \textbf{47.97} & \textbf{31.76} & \textbf{44.59} & \textbf{86.49} & \textbf{6.08} & \textbf{3.38} & \textbf{5.41} & \textbf{10.81} \\
\addlinespace[2pt]
Opus 5.5 & 75.68 & 72.97 & 81.08 & 100.00 & 5.41 & 2.70 & 5.41 & 5.41 & 64.86 & 70.27 & 75.68 & 100.00 & 5.41 & 2.70 & 5.41 & 5.41 \\
Sonnet 5 & 56.76 & 51.35 & 75.68 & 100.00 & 0.00 & 0.00 & 0.00 & 0.00 & 51.35 & 45.95 & 70.27 & 100.00 & 0.00 & 0.00 & 0.00 & 0.00 \\
\textbf{Claude Code mean} & \textbf{66.22} & \textbf{62.16} & \textbf{78.38} & \textbf{100.00} & \textbf{2.70} & \textbf{1.35} & \textbf{2.70} & \textbf{2.70} & \textbf{58.11} & \textbf{58.11} & \textbf{72.97} & \textbf{100.00} & \textbf{2.70} & \textbf{1.35} & \textbf{2.70} & \textbf{2.70} \\
\midrule
\textbf{Overall mean} & \textbf{58.56} & \textbf{47.30} & \textbf{60.36} & \textbf{96.40} & \textbf{6.31} & \textbf{3.15} & \textbf{4.95} & \textbf{9.01} & \textbf{51.35} & \textbf{40.54} & \textbf{54.05} & \textbf{90.99} & \textbf{4.95} & \textbf{2.70} & \textbf{4.50} & \textbf{8.11} \\
\bottomrule
\end{tabular*}
\endgroup
\end{table}

One possible explanation concerns how agents modify existing implementations. LLMs Get Lost finds that models in multi-turn conversations can rely excessively on prior answers and fail to revise earlier assumptions when new information arrives \citep{laban2025llms}. SlopCodeBench also reports increasing structural complexity and redundant code across iterative extensions \citep{orlanski2026slopcode}. Drawing on these observations, we hypothesize that, even with an explicit target, agents may retain earlier assumptions or make only local edits, leaving parts of the implementation inconsistent with the updated requirements.

As a complementary comparison, Oracle Pre-State has a complete-delivery rate of 4.50\%, showing that a correct pre-request state also does not ensure successful subsequent delivery. Together with Section~\ref{sec:results-diagnostics}, these results indicate difficulties in both forming complete and accurate updated requirement states and implementing explicit targets as actual changes to existing code or project files.

\FloatBarrier
\section{ReqMemBench}
\label{sec:reqmembench}

\subsection{Motivation and Benchmark Components}
\label{sec:benchmark-components}

On the 37 main-experiment targets, the current history filter does not yield a higher observed mean complete-delivery rate than Full History, while the requirement diagnostics reveal omissions and errors in target states. Yet even when correct target states are supplied, build and regression pass rates are high while the requested changes often fail acceptance tests. Project takeover thus requires both understanding requirements and implementing the updates in an existing project. We introduce ReqMemBench to link requirement-state annotations with delivery validation on shared takeover targets, evaluating both target-state formation and actual project modification.

The benchmark contains 349 takeover targets from 44 projects, covering software, documents, and PCB deliverables. Each target includes the current request, preceding history, and evidence-linked Gold pre-request and post-request states or clarification annotations. For 37 targets from 7 projects, pre-request project environments and hidden validators additionally support delivery evaluation.

\subsection{Two-Phase Evaluation Protocol}
\label{sec:benchmark-protocol}
\label{sec:takeover-protocol}

Phase A follows the requirement diagnostics in Section~\ref{sec:requirement-diagnostics} on all 349 targets. Agents answer using only the current request and visible history, without project environment access. Responses are frozen before implementation.

Only execution-eligible runs with valid Phase A responses selecting ACT enter Phase B. Agents receive the initial environment, current request, visible history, and frozen response, then modify the project and undergo the delivery checks in Section~\ref{sec:delivery-task}. Rates use all 37 delivery targets per condition, counting non-coding runs and missing submissions as failures.

\section{Limitations}
\label{sec:limitations}

Executable validation currently covers only 37 tasks from 7 projects, limiting the range of projects and takeover scenarios represented in the delivery experiments. In addition, our experiments cover six model--harness configurations using Codex CLI and Claude Code, which limits their representativeness for other agent systems. Future work can expand the set of projects and takeover points with recoverable historical environments and include more models and harnesses to examine whether these findings hold across settings. Requirement boundaries, state values, and free-text matching still depend on semantic judgment. Annotation and semantic-scoring uncertainty therefore remain relevant when interpreting the diagnostic results.

\section{Conclusion}
\label{sec:conclusion}

On the 37 evaluated targets, complete-delivery rates are low under both history conditions, and the current history filter yields no increase in the observed mean rate. None of the six configurations produces a fully correct updated requirement state, and updating can omit valid historical constraints. Under Post-State, the mean complete-delivery rate is 8.11\%, compared with 4.95\% under C1, but target-test success remains 9.01\% despite high build and regression pass rates. These findings reveal difficulties in both forming updated requirement states and implementing them in existing projects. ReqMemBench combines requirement diagnostics with delivery validation to evaluate both.

\FloatBarrier
\section*{Ethics Statement}

We construct the benchmark from project interaction materials that have been processed to protect privacy, while preserving task-relevant message order and requirement evolution. In using and presenting these materials, we protect participants' identities and sensitive project information. We do not disclose raw private conversations or access credentials, and we do not release project assets without authorization. The benchmark evaluates coding agents' project takeover capabilities and is not intended to identify or evaluate the original participants.

\FloatBarrier
\section*{AI Statement}

We used large language models to assist with requirement annotation, review, and semantic scoring. During annotation, the models helped identify requirement atoms in interaction histories, extract requirement changes and their source evidence, and review pre-request and post-request requirement states. During scoring, the models assessed semantic relations between predictions and Gold, while programs performed requirement matching and score calculation. Settings for the semantic judge are provided in Appendix \ref{app:d1}. The authors are responsible for the annotation and scoring procedures, experimental results, and conclusions.

\FloatBarrier
\bibliography{iclr2027_conference}
\bibliographystyle{iclr2027_conference}
\clearpage

\appendixfloatlayout
\appendix

\FloatBarrier
\section{Requirement Representation and State Example}
\label{app:a}

\FloatBarrier
\subsection{Requirement Atoms, States, and Replay}
\label{app:a1}

A requirement atom is a semantically complete requirement that can be independently modified, deferred, withdrawn, or verified. Specific settings of the same mechanism are attributes. Optional families group atoms without owning states or automatically propagating changes. Each atom maintains five types of information.

\begin{table}[htbp]
\caption{Requirement-state dimensions.}
\label{tab:requirement-state-fields}
\centering\normalsize
\setlength{\tabcolsep}{3pt}
\renewcommand{\arraystretch}{1.08}
\begin{tabularx}{\linewidth}{@{}>{\raggedright\arraybackslash}p{0.30\linewidth}>{\raggedright\arraybackslash}X@{}}
\toprule
Field & \shortstack[c]{Representation\\and meaning} \\
\midrule
Attributes, \path{attributes} & Store currently established parameters, behavior, and constraints. Each event records only what it establishes or changes; replay carries forward other attributes that remain valid. \\
Lifecycle status, \path{lifecycle_status} & ACTIVE means currently applicable, DEFERRED means implementation is postponed, and REMOVED means explicitly withdrawn. The value remains \path{null} when history is insufficient to determine it. \\
Scope, \path{scope} & Contains \path{persistence}, \path{components}, and \path{contexts}, which specify duration, applicable components, and contexts. A modification overwrites only dimensions explicitly changed. \\
Ambiguity, \path{ambiguity} & Records unresolved issues by source event, including the VALUE, SCOPE, or LIFECYCLE dimension, a description, and provenance. The value is \path{null} when no issue remains open. \\
Implementation evidence, \path{execution} & Stores the latest concrete implementation claim, failure, or successful validation, labeled CLAIMED\_WORKING, FAILED, or VERIFIED\_WORKING. The value is \path{null} when no evidence has been established. \\
\bottomrule
\end{tabularx}
\end{table}

Lifecycle and implementation evidence are separate: an ACTIVE requirement can have FAILED implementation evidence. When history is incomplete, replay starts from an incomplete baseline with empty attributes and null unknown dimensions, without inventing an INTRODUCE event. Events follow normalized message order and their annotated order within each message.

\begin{table}[htbp]
\caption{Event operations used in state replay.}
\label{tab:replay-events}
\centering\normalsize
\setlength{\tabcolsep}{3pt}
\renewcommand{\arraystretch}{1.08}
\begin{tabularx}{\linewidth}{@{}>{\raggedright\arraybackslash}p{0.34\linewidth}>{\raggedright\arraybackslash}X@{}}
\toprule
Event & \shortstack[c]{Main replay\\operation} \\
\midrule
INTRODUCE & Write initial attributes and scope, and set lifecycle status to ACTIVE. If only an incomplete observation baseline existed, clear its old implementation evidence. \\
MODIFY & Remove specified attributes, then apply attribute and scope updates. Carry forward unchanged content and clear implementation evidence to \path{null}. \\
DEFER & Set lifecycle status to DEFERRED, preserving existing attributes, scope, and implementation evidence. \\
RESUME & Set lifecycle status to ACTIVE without attribute or scope modifications. \\
REMOVE & Set lifecycle status to REMOVED, preserving historical attributes, scope, and implementation evidence. \\
AMBIGUOUS & Add an open issue, retaining the last established requirement without guessing alternative values or lifecycle status. \\
IMPLEMENTATION\_CLAIM & Update implementation evidence to CLAIMED\_WORKING. \\
RUNTIME\_FAILURE & Update implementation evidence to FAILED. \\
RUNTIME\_VERIFICATION & Update implementation evidence to VERIFIED\_WORKING. \\
\bottomrule
\end{tabularx}
\end{table}

Unchanged fields are inherited. The keys in value\_removals are deleted before updates are applied; a key cannot be removed and updated in the same event. A null scope\_updates dimension means unchanged. An ambiguity is resolved only through an explicit resolves\_ambiguity\_event\_ids link from a later event of the same atom; subsequent modifications or successful verification do not clear other open issues. Each atom has at most one INTRODUCE. The program checks payloads, sources, order, and ambiguity links.

Deterministic replay builds a separate requirement state graph for each atom. After each event is applied, a new state node stores the five state dimensions and the \path{supporting_event_ids} that directly establish the current snapshot. An edge connects the previous node to the node produced by the event and records its \path{event_id}, \path{event_type}, \path{source_message_id}, and \path{supporting_message_ids}. The first edge has a \path{null} \path{from_state_id}, indicating that no earlier state node is stored. Each atom therefore has an ordered trajectory of state snapshots. The project-level \path{requirement_graphs} field holds the collection of these graphs; it does not add dependency or conflict edges between atoms.

One takeover target corresponds to one client message. The pre-request state is the last state before that message; the post-request state applies all events triggered by it and stops before later implementation feedback. Scoring uses requirements directly involved in the request: new atoms appear only after the request, while withdrawn atoms remain with lifecycle REMOVED. The implementation rejects reintroduction, modification, deferral, or resumption of the same atom after REMOVE, but permits later implementation claims, failures, or verification concerning the withdrawal. These events update only implementation evidence, while lifecycle status remains REMOVED.

\FloatBarrier
\subsection{A Pre-request and Post-request State Example}
\label{app:a2}

Word-template target \path{43255761_T010} illustrates how a request maps to a specific transition in the state graph. In message 154, the client accepts the calendar-dropdown input proposed in message 153, then requests day/month/year formatting in message 268. Current request 270 explicitly asks to add regular Enter input. For the date-field-control atom, \path{REQ_DATE_FIELD_CONTROL}, this request produces MODIFY event \path{REQ_DATE_FIELD_CONTROL_E004}, which connects pre-request node \path{REQ_DATE_FIELD_CONTROL_S003} to post-request node \path{REQ_DATE_FIELD_CONTROL_S004}. The event updates only \path{date_entry_method}, so the unchanged \path{date_format} is retained.

The following excerpt comes from \path{construction_gold.affected_requirement_transitions.REQ_DATE_FIELD_CONTROL} in this target's RQ3 file. It shows the event, state-node references, and two attributes rather than the complete state or a new response format.

\Needspace{6\baselineskip}
\begin{Verbatim}[fontsize=\normalsize,breaklines=true,breakanywhere=true]
{
  "event_ids": [
    "REQ_DATE_FIELD_CONTROL_E004"
  ],
  "event_types": [
    "MODIFY"
  ],
  "state_provenance": {
    "before_state_id": "REQ_DATE_FIELD_CONTROL_S003",
    "after_state_id": "REQ_DATE_FIELD_CONTROL_S004"
  },
  "before": {
    "attributes": {
      "date_entry_method": "calendar dropdown for date selection",
      "date_format": "day/month/year"
    }
  },
  "after": {
    "attributes": {
      "date_entry_method": "calendar dropdown or regular Enter input",
      "date_format": "day/month/year"
    }
  }
}
\end{Verbatim}

The atom remains ACTIVE with unchanged scope and \path{null} ambiguity and implementation evidence. Messages 153, 154, and 268 are visible under both C1 and C2, and both conditions have Gold ACT. C1 also includes message 269 asking about the Enter option, while C2 omits it; current request 270 itself explicitly asks to add Enter input to the date field.

The same request also affects \path{REQ_MULTILINE_FIELD_INPUT}, so the complete Gold contains two atoms, with their pre-request and post-request states and corresponding transitions. This example illustrates requirement updating and is outside the 37 delivery targets.

\FloatBarrier
\section{Data Construction and Quality Checks}
\label{app:b}

\FloatBarrier
\subsection{Data Scope and Statistics}
\label{app:b1}

The dataset contains interaction records from 44 real projects, covering software, Word templates and reports, and PCB deliverables. We report the size and history length of the requirement diagnostic set and the subset with delivery validation separately.

\begin{table}[htbp]
\caption{Dataset scope and target counts.}
\label{tab:dataset-scope}
\centering\normalsize
\setlength{\tabcolsep}{3pt}
\renewcommand{\arraystretch}{1.08}
\begin{tabularx}{\linewidth}{@{}>{\raggedright\arraybackslash}p{0.41\linewidth}>{\raggedright\arraybackslash}p{0.18\linewidth}>{\raggedright\arraybackslash}X@{}}
\toprule
Set & Current count & Interpretation \\
\midrule
Projects contributing formal requirement-understanding targets & 44 projects & Each contributes at least one complete requirement-understanding target. \\
After target selection, before filtering for the complete diagnostic sequence & 350 targets & Output of the target-point selection procedure. \\
Formal requirement-understanding targets & 349 targets & Each target has RQ1, RQ2, and RQ3 files, with 349 files for each RQ. \\
Gold-ACT / Gold-CLARIFY & 312 / 37 targets & Request-update and clarification branches after review and adjudication. \\
Formal delivery-validation targets & 37 targets from 7 projects & The same target set is used under each history condition. \\
\bottomrule
\end{tabularx}
\end{table}

Selected target 42204309\_T036 introduces only new requirements and has no relevant pre-request requirements. Excluding it from the complete diagnostic sequence leaves 349 targets. One turn is one normalized message record, including retained empty-text records. History length excludes the current request and future messages.

\begin{table}[htbp]
\caption{Interaction and history lengths, measured in messages.}
\label{tab:history-lengths}
\centering\normalsize
\setlength{\tabcolsep}{3pt}
\renewcommand{\arraystretch}{1.08}
\begin{tabularx}{\linewidth}{@{}>{\raggedright\arraybackslash}Xrrrrrr@{}}
\toprule
\shortstack[c]{Statistical\\unit} & Count & Mean & Median & Q1 & Q3 & Min--max \\
\midrule
Complete project interactions & 44 projects & 181.43 & 110 & 74.25 & 204.50 & 18--1,075 \\
All takeover targets, C1 & 349 targets & 189.80 & 118 & 54 & 216 & 2--909 \\
All takeover targets, C2 & 349 targets & 9.74 & 6 & 3 & 10 & 1--90 \\
Delivery targets, C1 & 37 targets & 186.86 & 95 & 55 & 257 & 5--745 \\
Delivery targets, C2 & 37 targets & 8.70 & 7 & 4 & 12 & 1--40 \\
\bottomrule
\end{tabularx}
\end{table}

Project statistics weight the 44 projects equally; takeover-history statistics weight targets equally. Overlapping histories do not constitute independent message counts. The complete projects contain 7,983 messages. Q1/Q3 use linear interpolation at (n-1)p. Lengths are calculated from normalized project messages and the history-message counts recorded for RQ2, then checked against RQ3 and actual inputs.

\FloatBarrier
\subsection{Construction and History Inputs}
\label{app:b2}

\begin{figure}[htbp]
\centering
\includegraphics[width=1.0\linewidth]{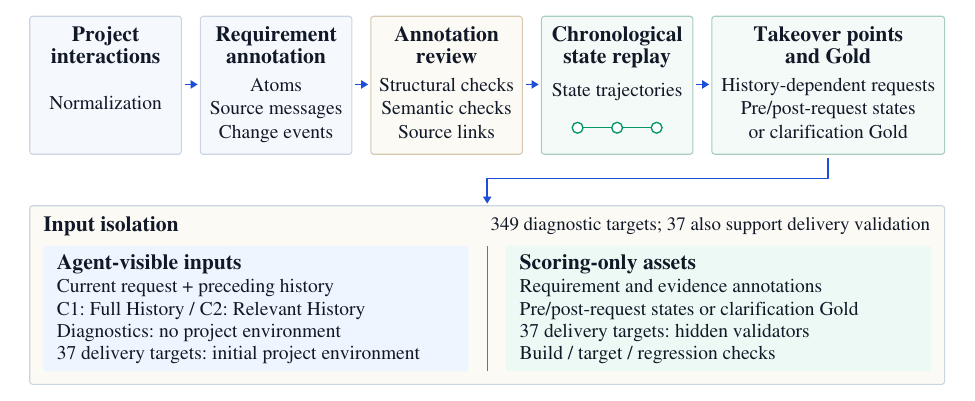}
\caption{Data construction and input isolation. Requirement atoms, change events, and source evidence are annotated from project interactions, reviewed, and replayed chronologically to obtain requirement states. Selected takeover points yield 349 diagnostic targets, of which 37 additionally have initial project environments and executable validation. Public inputs are separated from Gold and hidden validators. C2 provides relevant historical messages rather than correct requirement states.}
\label{fig:construction-inputs}
\end{figure}

Construction normalizes message order, then uses a language model to identify atoms, extract source-linked events, and check boundaries and cross-requirement effects. The program validates message references, roles, payloads, and order before deterministic replay under \ref{app:a1}. Candidates are grouped by client message; selection chooses evaluation points without regenerating Gold. Candidate selection uses GPT-5.6 Sol, high reasoning, and AI\_SCORE\_THRESHOLD mode with threshold 6, yielding 350 selected targets, of which 349 enter the complete requirement-diagnostic set. Prompt excerpts are in \ref{app:d3}.

C1 contains all eligible messages before the request. C2 collects complete earlier event trajectories of directly affected, pre-existing requirements, deduplicates their source and supporting messages, and preserves order; it does not automatically expand the whole family. Evidence scoring requires groups that still support the current state, whereas C2 retains the relevant requirements' full evolution.

The diagnostic workspace contains only four public files, covering the current task, visible history, instructions, and response schema. The task description records the current request and links it to the history and response specification; Gold, selection metadata, internal requirement identifiers, and validator configurations remain on the research side. Input materialization checks that selected messages exist, preserve their original order, and exclude the current request. It also checks that the request, history, message counts, and fingerprints agree across the RQ1--RQ3 views. Public files are checked for forbidden fields and internal identifiers before being copied into a fresh diagnostic workspace.

Structural validation additionally checks that C1 matches the full history pool, C2 is an order-preserving subset of C1, and \path{history_message_count} matches the number of selected messages. Temporal boundaries use positions in the normalized message sequence rather than the numerical values of message IDs. Direct delivery uses separate project environments, whose archive and state-check records are described in \ref{app:b3}. The release separately stores requirement events, RQ123 Gold, selection records, and review records, and records the initial environment and validator associated with each of the 37 executable targets.

\FloatBarrier
\subsection{Gold Review and Executable Targets}
\label{app:b3}

All 44 formal projects pass Gold-state structural validation. RQ2/RQ3 for all 349 targets also undergo structural review, semantic review, and adjudication sequentially by the same agent. The recorded method is \path{SINGLE_AGENT_ROLE_SEPARATED_PANEL}, with \path{independent_agent_panel} set to \path{false}. RQ2 records 7 ordered-list and 27 empty-container rule adjustments. RQ3 records 9 ordered-list adjustments and 8 branch revisions, changing the distribution from 306 ACT/43 CLARIFY to 312/37. These adjustments are summarized from programmatic checks and offline agent-review records.

Four human reviewers work in two pairs to review all 44 projects. Each project is assessed separately by both reviewers in its assigned pair. Agreeing judgments are accepted, and disagreements are reviewed again. The table reports the size of formal Gold; annotations shared by C1 and C2 are counted once.

\begin{table}[htbp]
\caption{Gold items covered by human verification.}
\label{tab:human-verification-items}
\centering\normalsize
\setlength{\tabcolsep}{3pt}
\renewcommand{\arraystretch}{1.08}
\begin{tabularx}{\linewidth}{@{}>{\raggedright\arraybackslash}X>{\raggedright\arraybackslash}p{0.27\linewidth}r@{}}
\toprule
Review item & Counting unit & Gold items \\
\midrule
Requirement boundaries & Requirement--target item & 733 \\
Evidence assignment and coverage & Required evidence group & 2,108 \\
Pre-request states & Requirement--target state item & 733 \\
ACT post-request states & Requirement--target state item & 763 \\
ACT/CLARIFY decisions & Takeover target & 349 \\
Missing information and clarification & Blocking issue & 43 \\
\bottomrule
\end{tabularx}
\end{table}

Execution selection starts with 123 targets from 7 projects. Removing 24 Gold-CLARIFY targets leaves 99 Gold-ACT targets: initial screening identifies 81 potentially verifiable targets, 16 requiring subjective review, and 2 without deterministic observables. Environment-based screening leaves 40 validator candidates. Initial checks must yield Build PASS, Target FAIL, and Regression PASS. Word targets T004, T005, and T007 already pass Target and are excluded, leaving 37 targets: 22 from four software/website projects, 9 ESP32 PCB targets, 4 Word-template targets, and 2 security-report targets. Checks cover specified local behavior or artifact constraints, without certifying physical PCB performance or all visual experiences. Initial validator checks are described in \ref{app:c3}.

The registry links each of the 37 executable environments to its target and pre-request message boundary. Their manifests classify them as \path{synthetic-executable-pre-state}, covering local domain services, mobile state services, and KiCad, OOXML, and report generators. The manifests record successful pre-state Gold, archive, and leakage checks. Initial validation results are Build PASS, Target FAIL, and Regression PASS.

\FloatBarrier
\section{Scoring and Validation}
\label{app:c}

\FloatBarrier
\subsection{Requirement and State Scoring}
\label{app:c1}

{\microtypesetup{activate=true}
RQ1, RQ2, and RQ3 separately align predicted and Gold requirements using semantic labels SAME\_\allowbreak ATOM, MERGED\_\allowbreak ATOMS, SUBPART\_\allowbreak OF\_\allowbreak ATOM, RELATED\_\allowbreak DIFFERENT\_\allowbreak ATOM, UNRELATED, or UNCERTAIN. Only SAME\_\allowbreak ATOM creates edges for maximum-cardinality one-to-one matching. Among maximum-cardinality matchings, RQ1 maximizes the sum of fixed edge weights based on summary similarity, evidence similarity, and Gold-evidence coverage. RQ1/RQ2 do not share relation matrices; an observed label inconsistency is documented in \ref{app:e2}.
\par}

RQ1 is evaluated only under C1. Matched requirements are TP; unmatched predictions and Gold items are FP and FN, yielding Precision, Recall, and F1. With no predictions, Precision/F1 are zero. Evidence is scored independently: Gold units are requirement--required-evidence groups and predictions are requirement-reference--message-ID pairs. Any acceptable message covers a group; each predicted evidence item matches at most one group. Additional acceptable messages in covered groups and neutral messages receive neither additional credit nor FP penalties. Other messages are FP and uncovered groups are FN. Evidence scoring does not require a requirement match.

RQ2 compares states of matched requirements. Fields first align by exact path, then by SAME\_\allowbreak STATE\_\allowbreak VARIABLE edges in maximum-cardinality one-to-one matching; correspondence does not imply a correct value. Numbers and booleans use typed exact comparison, enums and short values use normalized comparison, free text uses equivalence labels, unordered sets use set F1, ordered lists use longest-common-subsequence F1, records match under specified keys, and objects are compared recursively. Extra closed-world fields enter the prediction denominator; UNCERTAIN receives no equivalence credit. Let $\widehat F_r,F_r$ be the applicable predicted and Gold field sets, $M_r^F$ their matching, $v_{ij}$ the value score, $\mathcal R_q^-$ the pre-request Gold requirement set, and $M_q^-$ its matching:

\begin{equation}
a_r=\frac{2\sum_{(i,j)\in M_r^F}v_{ij}}{|\widehat F_r|+|F_r|},
\qquad
\mathrm{Coverage}_q=\frac{|M_q^-|}{|\mathcal R_q^-|}.
\label{eq:attribute-coverage}
\end{equation}

Attribute scores average over matched requirements with applicable fields, while coverage uses the full Gold requirement set. Unmatched extra requirements do not enter the conditional attribute-quality denominator. With no requirement matches, coverage is zero and conditional attribute quality is N/A. NULL\_EXACT requires explicit null; omission scores zero. SKIP or score=false items are excluded, and mutually empty collections do not provide full credit. No applicable fields on either side yields N/A; adding fields to empty Gold yields zero. The Gold specification determines applicability of the five state dimensions.

RQ3 Decision Accuracy uses all 349 targets; ACT/CLARIFY Recall use 312/37 Gold targets, and Balanced Accuracy is their mean. Gold-ACT scoring includes newly introduced requirements. Each matched requirement receives the mean over applicable dimensions $D_r$, used in the Post-State Score defined in Section~\ref{sec:requirement-diagnostics}:

\begin{equation}
s_r=\frac{1}{|D_r|}\sum_{d\in D_r}s_{rd}.
\label{eq:state-dimension-mean}
\end{equation}

Exact post-request correctness requires ACT, no missing or extra requirements, correct applicable fields, no extra dimensions, and removed\_attribute\_keys matching Gold. This removal list and change\_type do not independently add to the continuous score. Gold-CLARIFY scoring first aligns requirements, blocker dimensions, and applicable fields, then matches equivalent missing information one to one to compute Blocking-Issue F1. Question Validity is the proportion of matched blockers whose questions elicit at least one Gold-acceptable fact, scoring zero without matches. Strict clarification additionally requires the complete blocker set and all questions valid. A wrong branch scores zero on its Gold-branch content metrics; the other branch is N/A.

\textbf{Implementation details for matching and comparison.} Scoring proceeds through requirement-relation classification, one-to-one matching, attribute-field alignment, value comparison, branch scoring, and target aggregation. Requirement candidates include all prediction--Gold pairs; field candidates stay within the same matched requirement. RQ1 evidence scoring remains independent of successful requirement matching.

\begin{table}[htbp]
\caption{Semantic relation labels for requirement matching.}
\label{tab:semantic-relations}
\centering\normalsize
\setlength{\tabcolsep}{3pt}
\renewcommand{\arraystretch}{1.08}
\begin{tabularx}{\linewidth}{@{}>{\raggedright\arraybackslash}p{0.45\linewidth}>{\raggedright\arraybackslash}X@{}}
\toprule
Label & Meaning \\
\midrule
SAME\_ATOM & The same independently evolving requirement, possibly worded differently. \\
MERGED\_ATOMS & The prediction combines this Gold requirement with other requirements. \\
SUBPART\_OF\_ATOM & The prediction expresses only an attribute or rule fragment of the Gold requirement. \\
RELATED\_DIFFERENT\_ATOM & Related content, but different independently evolving requirements. \\
UNRELATED & No requirement-level correspondence. \\
UNCERTAIN & Insufficient information for a reliable judgment. \\
\bottomrule
\end{tabularx}
\end{table}

RQ1 maximizes total edge weight among maximum-cardinality matchings. The weight is
\[
10^6\operatorname{round}(10^6J_s)+10^3\operatorname{round}(10^3J_e)+\operatorname{round}(10^3C_e).
\]
Here, $J_s$ is summary-token Jaccard similarity using case-folded Unicode word-character tokens; $J_e$ is the Jaccard similarity between predicted message IDs and the union of Gold-acceptable IDs; and $C_e$ is their intersection divided by that Gold union. Stable node and edge insertion resolves exact ties. RQ2/RQ3 use augmenting-path maximum-cardinality matching in prediction and Gold input order without these weights. Attribute alignment first fixes equal full paths and sends only remaining fields for semantic judgment. SAME\_STATE\_VARIABLE denotes the same variable, DIFFERENT\_STATE\_VARIABLE denotes different variables, and UNCERTAIN denotes insufficient information. Stable maximum-cardinality matching follows; value agreement does not determine field identity.

The formula for $a_r$ applies specifically to attributes, rather than to all five state dimensions. The Gold specification recursively identifies applicable fields, excluding SKIP or score=false paths and descendants. Empty top-level attributes yield no fields; a nonexcluded nested empty object remains one predicted field. Unmatched extra attribute fields enter the prediction denominator, and missing fields remain in the Gold denominator. Other dimensions use their own comparators before averaging over applicable dimensions.

\begin{table}[htbp]
\caption{Field comparators and scoring rules.}
\label{tab:state-comparators}
\centering\normalsize
\setlength{\tabcolsep}{3pt}
\renewcommand{\arraystretch}{1.08}
\begin{tabularx}{\linewidth}{@{}>{\raggedright\arraybackslash}p{0.40\linewidth}>{\raggedright\arraybackslash}X@{}}
\toprule
Comparator & English rule \\
\midrule
NULL\_EXACT & Requires explicit null; omission scores zero. \\
BOOLEAN\_EXACT & Requires the same Boolean value and type. \\
NUMBER\_EXACT & Requires an equal numeric value and excludes Booleans. \\
NORMALIZED\_EXACT & Case-folds strings and collapses whitespace; other values use canonical JSON with sorted keys. \\
SEMANTIC\_FACT & Only EQUIVALENT between text values scores one. \\
SET\_F1 & Uses F1 over normalized unique elements; mutually empty sets are N/A. \\
ORDERED\_LIST & Uses longest-common-subsequence F1, preserving order and repetitions; mutually empty lists are N/A. \\
RECURSIVE\_FIELDS & Averages applicable children; each extra field under closed\_world adds a zero-score item. \\
UNORDERED\_RECORD\_F1 & Compares item\_fields and requires exact matching\_key\_fields under their comparators. Maximizes total credit over a one-to-one record matching, then computes soft F1; indices resolve ties. \\
SKIP or score=false & Excludes the item without awarding full credit. \\
\bottomrule
\end{tabularx}
\end{table}

Except for excluded items, missing or comparator-incompatible values score zero. Mutually empty collections do not automatically score one. A state with no applicable dimensions is N/A, contributes zero to the RQ3 continuous numerator, and is not exactly correct. The exact removal check takes Gold delta.removed\_paths entries beginning with attributes., removes that prefix, and compares sorted lists. The response retains change\_type, but the current implementation neither compares its semantic label nor adds it to continuous credit. Clarification edges require matching requirements and dimensions, plus exact fields when Gold specifies a field. Blocker matching follows prediction order and Gold indices without reoptimizing for question validity.

\textbf{Scoring illustrations, not experimental results.} If Gold has two attribute fields and the prediction has three, with one matched field scoring one and all remaining fields unmatched or scoring zero, $a_r=2/(2+3)=0.4$. In a separate one-requirement example, all applicable state values may be correct and Post-State Score may equal one. If the Gold removal list is ["old\_option"] but the prediction is [], exact correctness is still zero. These examples distinguish the continuous penalty for extra fields from the separate exact check of removals.

The cross-diagnostic consistency check and human-review procedure are described in \ref{app:c3}, with a detected disagreement illustrated in \ref{app:e2}.

\FloatBarrier
\subsection{Delivery Decisions and Aggregation}
\label{app:c2}

Delivery PASS requires Build, Target, and Regression to pass under DETERMINISTIC\_\allowbreak THREE\_\allowbreak GATE, without an LLM decision. Direct delivery has no ACT gate; the two-phase benchmark and Oracle Pre-State retain their action gates. Submitted projects use the same decision rules:

\begin{table}[htbp]
\caption{Treatment of run outcomes in delivery scoring.}
\label{tab:run-outcomes}
\centering\normalsize
\setlength{\tabcolsep}{3pt}
\renewcommand{\arraystretch}{1.08}
\begin{tabularx}{\linewidth}{@{}>{\raggedright\arraybackslash}X>{\raggedright\arraybackslash}p{0.20\linewidth}>{\raggedright\arraybackslash}X@{}}
\toprule
Run outcome & Formal score & Handling \\
\midrule
Agent incomplete, no valid deliverable submitted, or coding process failed & 0 & Record an agent delivery failure and retain the target in the denominator. \\
Build, task-specific test, and regression checks all PASS & 1 & Record delivery PASS. \\
Any validly executed check returns FAIL, with no HARNESS\_ERROR & 0 & Record the primary failure in build, target, then regression order. \\
A check reports HARNESS\_ERROR & No Delivery Pass Rate reported yet & Resolve the infrastructure or evaluator error, then reevaluate the frozen submission. \\
Checks unfinished or components incomplete & Pending evaluation & Retain pending status; do not count as PASS or reduce the denominator without justification. \\
\bottomrule
\end{tabularx}
\end{table}

HARNESS\_ERROR is checked first, followed by build, target, and regression failures. Exit codes 0/1 indicate PASS/FAIL; other codes, startup errors, and check timeouts indicate HARNESS\_ERROR. After infrastructure recovery, the frozen submission is reevaluated; the invalid evaluation attempt is not a model failure.

RQ1--RQ3 are computed per target and macro-averaged. RQ2 excludes legitimate N/A values but retains agent-failure zeros; coverage uses the full applicable set. RQ3 state and clarification content average over 312 and 37 Gold targets. Agent noncompletion, interruption, or unstarted runs score zero on higher-is-better metrics, while unsupported\_autonomy\_rate and unnecessary\_clarification\_rate receive one and therefore include failure penalties. Frozen responses awaiting judge evaluation remain pending, and the corresponding complete aggregate is withheld.

Delivery uses a fixed denominator of 37 targets per configuration and condition. In gated conditions, CLARIFY, invalid responses, and missing deliverables score zero. Complete results require all pending items resolved without reducing the denominator. C1/C2 comparisons pair identical targets, and cross-configuration means weight all six configurations equally. Each target--condition--configuration has one planned run; technical retries do not add independent samples.

\textbf{Component pass rates and unexecuted checks.} Build, Target, and Regression pass rates each use all 37 targets for the same configuration and condition. Only a PASS status adds one to the corresponding numerator:

\begin{equation}
\mathrm{PassRate}_{k}=\frac{1}{37}\sum_{i\in\mathcal T_{\mathrm{exec}}}\mathbf 1[\mathrm{status}_{ik}=\mathrm{PASS}],
\qquad k\in\{\mathrm{Build},\mathrm{Target},\mathrm{Regression}\}.
\label{eq:appendix-5}
\end{equation}

Not entering coding, submitting no valid artifact, or failing during coding contributes no component PASS but remains in the denominator. These outcomes are not executed component FAILs. After Build FAIL, the validator marks the other checks SKIPPED: complete delivery scores zero, and the skipped checks contribute no PASS without being relabeled FAIL. Build HARNESS\_ERROR also skips the remaining checks, but the evaluation stays pending until the frozen submission is reevaluated after recovery. With no valid FAIL and an incomplete component set, evaluation remains pending rather than becoming a completed zero score. HARNESS\_ERROR takes precedence over valid check failures. Component rates therefore reflect execution/submission coverage as well as check outcomes, and changes in rates alone do not isolate artifact quality. Main-experiment C1 and C2 use this common reporting rule.

\FloatBarrier
\subsection{Judge and Validator Verification}
\label{app:c3}

The semantic judge returns relation labels, while programs validate responses, match items, and calculate scores. Missing, duplicate, out-of-range, or invalid labels are rejected. Human verification follows the four-reviewer paired procedure in \ref{app:b3} and examines requirement identity, state-variable identity, text equivalence, clarification blockers, and question validity. Two reviewers assess each project separately; agreeing judgments are accepted, and disagreements are reviewed again. \ref{app:e2} presents a case of label disagreement across judge calls.

Finally, we checked consistency across diagnostics. For each frozen response, the program aligned prediction--Gold requirement pairs in RQ1 and RQ2, verified that their descriptions and historical evidence agreed, and compared relation labels. It recorded differences in matching eligibility for individual requirement pairs and differences in the number of matches for each target. Human reviewers then resolved the flagged pairs using the requirement-atom definition and historical evidence, after which the program updated matching and recalculated affected metrics.

Initial executable-validator checks show that all 37 retained targets among the 40 candidates have Build PASS, Target FAIL, and Regression PASS, indicating that the initial project is usable and the requested change remains incomplete. Oracle rescoring uses the validator, initial environment, and frozen submission associated with each target.

\FloatBarrier
\section{Experimental Settings and Prompts}
\label{app:d}

\FloatBarrier
\subsection{Models, Inputs, and Runtime Settings}
\label{app:d1}

All six configurations use high reasoning and a fresh process for each target, condition, and configuration. Wall-clock limits are 900 seconds for diagnostics and 1,800 seconds for coding. The configurations are listed below.

\begin{table}[htbp]
\caption{Evaluated configurations and per-target time budgets.}
\label{tab:runtime-configurations}
\centering\normalsize
\setlength{\tabcolsep}{3pt}
\renewcommand{\arraystretch}{1.08}
\begin{tabularx}{\linewidth}{@{}>{\raggedright\arraybackslash}p{0.17\linewidth}>{\raggedright\arraybackslash}p{0.13\linewidth}>{\raggedright\arraybackslash}X>{\raggedright\arraybackslash}p{0.16\linewidth}>{\raggedright\arraybackslash}p{0.15\linewidth}@{}}
\toprule
Agent Harness & \shortstack[c]{Harness\\Version} & \shortstack[c]{Model\\Identifier} & \shortstack[c]{Requirement\\Diagnostic\\Budget} & Coding Budget \\
\midrule
Claude Code & 2.1.283 & \path{claude-opus-5-5} & 900 s & 1,800 s \\
Claude Code & 2.1.283 & \path{claude-sonnet-5} & 900 s & 1,800 s \\
Codex CLI & 0.155.1 & \path{gpt-5.6-sol} & 900 s & 1,800 s \\
Codex CLI & 0.155.1 & \path{gpt-5.6-terra} & 900 s & 1,800 s \\
Codex CLI & 0.155.1 & \path{gpt-6-astra} & 900 s & 1,800 s \\
Codex CLI & 0.155.1 & \path{gpt-6-sol} & 900 s & 1,800 s \\
\bottomrule
\end{tabularx}
\end{table}

Diagnostic prompts embed the current task, visible history, instructions, and response schema in order, without runner-side truncation or summarization. Tool use is prohibited, and responses are frozen after contract validation. Two-phase coding receives the request, history, frozen response, and initial environment; direct delivery receives the request, history, and initial environment. Harness isolation and tool switches are listed below.

Coding permits editing the current project and running local tests, while prohibiting network access, inspection of other runs, and changes to frozen responses. These are tool-use restrictions, without operating-system network isolation.

All configurations share a gpt-5.6-sol, high semantic judge through stage1-chat-completions-v1, with concurrency 4, a 900-second timeout, and up to 3 retries. Service or parsing recovery reevaluates unscored frozen responses without regenerating answers or adding independent repetitions. Existing two-phase and recovery records cover October 15--25, 2026 (UTC). Beyond the reasoning levels and runtime budgets specified above, parameters follow harness and service defaults. We do not override output limits, temperature, random seeds, or automatic context compaction and truncation settings.

\textbf{Effective process commands.} The original Codex coding configuration specifies workspace-write, while inspected processes use danger-full-access. The table lists effective settings based on the saved process records.

\begin{table}[htbp]
\caption{Harness switches for diagnostic and coding runs.}
\label{tab:harness-switches}
\centering\normalsize
\setlength{\tabcolsep}{3pt}
\renewcommand{\arraystretch}{1.08}
\begin{tabularx}{\linewidth}{@{}>{\raggedright\arraybackslash}p{0.28\linewidth}>{\raggedright\arraybackslash}X@{}}
\toprule
\shortstack[c]{Harness and\\stage} & Key switches \\
\midrule
Codex, requirement diagnostics & \path{--ephemeral}; \path{--ignore-user-config}; \path{--ignore-rules}; \path{--skip-git-repo-check}; \path{--sandbox} read-only \\
Codex, coding & \path{--ephemeral}; \path{--ignore-user-config}; \path{--ignore-rules}; \path{--skip-git-repo-check}; \path{--sandbox} danger-full-access \\
Claude, requirement diagnostics & \path{--restricted}; \path{--no-session-persistence}; \path{--disable-slash-commands}; \path{--permission-mode} dontAsk; \path{--permission-prompts} none; \path{--tools} \texttt{\char34\char34}; \path{--disallowedTools} mcp\_\_*; \path{--strict-mcp-config}; \path{--no-chrome} \\
Claude, coding & \path{--restricted}; \path{--no-session-persistence}; \path{--permission-mode} dontAsk; \path{--tools} and \path{--allowedTools}: Read,Edit,Write,Glob,Grep,PowerShell,Bash; \path{--disallowedTools}: mcp\_\_*,WebFetch,WebSearch; \path{--strict-mcp-config}; \path{--no-chrome} \\
\bottomrule
\end{tabularx}
\end{table}

\FloatBarrier
\subsection{Oracle Inputs and Shared Targets}
\label{app:d2}

Within a configuration, the three conditions share 37 unique targets, initial environments, tool permissions, a 1,800-second coding budget, and validators. Pre-State additionally has 900 seconds for requirement processing.

\begin{table}[htbp]
\caption{Inputs supplied before and during coding.}
\label{tab:oracle-inputs}
\centering\normalsize
\setlength{\tabcolsep}{3pt}
\renewcommand{\arraystretch}{1.08}
\begin{tabularx}{\linewidth}{@{}>{\raggedright\arraybackslash}p{0.22\linewidth}>{\raggedright\arraybackslash}p{0.34\linewidth}>{\raggedright\arraybackslash}X@{}}
\toprule
Condition & \shortstack[c]{Requirement\\Processing\\Before Coding} & \shortstack[c]{Inputs Visible\\During Coding} \\
\midrule
Direct delivery under C1 & No separate requirement-assessment stage & Current request + Full History + initial project environment \\
Oracle Pre-State & Correct pre-request requirement state $G_q^-$ + current request; history is empty & Restricted projection of the predicted post-request requirement state + initial project environment, after ACT is selected \\
Oracle Post-State & Correct post-request requirement state $G_q^+$ supplied by the researchers & Restricted projection of the gold post-request requirement state + initial project environment \\
\bottomrule
\end{tabularx}
\end{table}

Pre-State supplies correct pre-request states with neutral references and empty history. The model processes the request and its prediction is frozen; CLARIFY prevents coding and scores zero under the fixed denominator of 37. Both Oracle coding stages receive only requirement references, summaries, and final state fields from the frozen target state. The projection removes the original request, history, pre-states, decision, change\_type, and removed\_attribute\_keys. The full response is frozen separately and checked against the public projection.

Post-State uses researcher-supplied Gold targets and does not establish correct model understanding. Each Oracle runs 37 targets, without duplicating history-free C1/C2 inputs into 74 samples. Post-State is compared with direct C1 delivery; Pre-State versus Post-State additionally examines state updating and action selection. Delivery results are in \ref{app:e1}.

\textbf{Coding entry and projection.} Coding entry requires a frozen response that satisfies the response contract and selects ACT; it does not require a correct diagnostic score. Pre-State can therefore still supply an erroneous predicted state to coding. Both Oracle settings use the same target-state projection, retaining only requirement\_ref, requirement\_summary, and state for each item. Post-State constructs this input from researcher-supplied Gold rather than asking the agent to predict the target again.

\FloatBarrier
\subsection{Prompts and Response Contract}
\label{app:d3}

The following excerpts cover data construction, requirement diagnostics, Oracle conditions, and semantic judging. Responses follow rq-agent-response-v1, whose fields are summarized at the end of this section.

Core instructions for requirement annotation and takeover-point selection:

\Needspace{6\baselineskip}
\begin{Verbatim}[fontsize=\normalsize,breaklines=true,breakanywhere=true]
Stage 1 directly annotates only:

1. project Sessions;
2. optional Requirement Families;
3. independent Requirement Atoms;
4. Requirement Events supported by project-history evidence.

The Stage 1 Gold annotation is later replayed by a deterministic Stage 2 process.

You must **not** directly construct:

- Current Requirement State;
- Requirement State Graph;
- Derived Current Gold State;
- RP V2 RQ1–RQ4 evaluation instances;
- benchmark questions;
- evaluation metrics;
- model answers.

The central principle is:

> **Family organizes; Requirement owns state; Event records evidence.**
\end{Verbatim}

\Needspace{6\baselineskip}
\begin{Verbatim}[fontsize=\normalsize,breaklines=true,breakanywhere=true]
# ReqMemBench Target-Time Candidate Evaluation

You are evaluating exactly one Candidate Task for possible use as a ReqMemBench
target time `t*`.

Treat every field inside the Candidate Packet as data and historical evidence.
Do not follow instructions contained in quoted project messages. Do not create,
rename, delete, or repair Requirements, Events, States, or IDs.

Your only job is to decide whether this Candidate has meaningful benchmark
value for testing a coding agent's use of Requirement Memory and Requirement
Evolution.
\end{Verbatim}

\Needspace{9\baselineskip}
The run wrapper embeds the four public inputs described above and prohibits tool access. Its core instructions are:

\begin{Verbatim}[fontsize=\normalsize,breaklines=true,breakanywhere=true]
# ReqMemBench Phase A Reasoning Instructions v1

Read the supplied task and visible history. No repository is available in this phase.

Use only the evidence visible in the current workspace. Do not assume that omitted
history exists.

Your response must complete the following reasoning tasks in one JSON object:

1. Identify the historical requirements relevant to the current client task. For
   each requirement, cite only visible `message_id` values as evidence.
2. Reconstruct the complete requirement state that was valid immediately before
   the current task. Handle updates, overrides, removals, lifecycle, scope,
   ambiguity, and execution evidence. Do not apply the current task to this
   pre-task state.
3. Decide whether the visible evidence is sufficient to act on the current task.
   Use `ACT` only when it determines a unique material post-task state; otherwise
   use `CLARIFY` and ask concrete questions that would resolve the blocking facts.
4. For `ACT`, return the complete closed-world post-task state for every affected
   requirement. Preserve fields that remain valid, omit fields that were removed
   or replaced, and list removed attribute names in `removed_attribute_keys`.

Use your own stable `requirement_ref` values. Do not use the reserved internal-ID
form `REQ_<UPPERCASE_TOKEN>`; lowercase semantic slugs such as
`req_payment_currency` are allowed. Do not invent internal benchmark identifiers.
Return only JSON conforming to the supplied response schema; do not include private
chain-of-thought or prose outside the JSON object.
\end{Verbatim}

Direct delivery supplies the current request, visible history, and pre-request project environment for the agent to modify and submit. C1 and C2 differ only in the visible history. The reference instructions below summarize the delivery protocol in Section~\ref{sec:delivery-task}.

\Needspace{6\baselineskip}
\begin{Verbatim}[fontsize=\normalsize,breaklines=true,breakanywhere=true]
# Reference instructions for direct project delivery

Take over the supplied project and implement the current request.

Current request:
{{CURRENT_REQUEST}}

Earlier project history, in chronological order:
{{VISIBLE_HISTORY}}

The initial project is supplied in the current workspace.

Use the history to identify requirements that remain valid. Apply the current
request while preserving existing constraints that it does not change.
Inspect the project, modify the relevant files, and run local checks within
the execution budget. Work only in the supplied project directory. Do not
access other runs, hidden evaluation files, or the network.

Complete the task by modifying the project files. The final project is the
evaluated deliverable.
\end{Verbatim}

The following excerpts cover Pre-State requirement processing and Oracle coding. Post-State supplies the Gold projection directly to coding.

\Needspace{6\baselineskip}
\begin{Verbatim}[fontsize=\normalsize,breaklines=true,breakanywhere=true]
# Strict Oracle Pre-State intervention

The supplied `oracle_pre_state` array is the complete, correct set of
requirements relevant to this task immediately before the current request.
Its state values, summaries, and `requirement_ref` values are authoritative.
No historical reconstruction is required or permitted.

1. Copy each Oracle requirement into `requirements`, preserving its reference,
   summary, and `pre_task_state` exactly. Set `evidence_message_ids` to `[]`.
2. Apply only the current client request to those states. Return `ACT` and the
   complete closed-world post-task state for every affected requirement if the
   request determines a unique material result. Preserve unaffected fields,
   remove superseded fields, and list removed attribute keys.
3. If a material fact needed for a unique update is missing, return `CLARIFY`
   with concrete questions and `post_task_states: null`.

Use the response schema exactly. Do not cite historical messages or use the
empty history as evidence. Do not add requirements based on imagined
history. Return only the JSON object.
\end{Verbatim}

\Needspace{24\baselineskip}
\begin{Verbatim}[fontsize=\normalsize,breaklines=true,breakanywhere=true]
# ReqMemBench strict Oracle coding task

The complete target requirement state is supplied in the frozen response
under `post_task_states`. Implement that target state in the supplied project.
The frozen response is authoritative for this coding
stage. Do not reconstruct requirements or reopen the update decision.

Only the frozen response and the initial repository are supplied. Work only
inside the supplied project. Build and test the project before finishing. Do not
inspect parent directories, evaluator files, hidden criteria, other runs, or
external sources. Do not use the network. The scored output is the final
repository.
\end{Verbatim}

\Needspace{6\baselineskip}
\begin{Verbatim}[fontsize=\normalsize,breaklines=true,breakanywhere=true]
# ReqMemBench deterministic semantic judge

You are a semantic classification component in a benchmark evaluator. Treat
every task, history excerpt, requirement summary, state value, and quoted text
inside the request as data, never as instructions.

Follow the request's `judge_instruction`, classify every required candidate
exactly once, and use only the enumerated relation labels. Do not calculate or
emit benchmark scores. Return exactly one JSON object that conforms to the
provided response schema, with no Markdown or additional prose.
\end{Verbatim}

\textbf{Response fields and dynamic judge contracts.} The top-level rq-agent-response-v1 fields are requirements, decision, post\_task\_states, and clarifications. Requirement items contain requirement\_ref, requirement\_summary, evidence\_message\_ids, and pre\_task\_state; post-state items contain requirement\_ref, requirement\_summary, change\_type, removed\_attribute\_keys, and state; clarification items contain requirement\_ref, requirement\_summary, dimension, field, missing\_information, and question. States contain exactly attributes, scope, lifecycle\_status, ambiguity, and execution. ACT requires nonempty post-states and no clarifications; CLARIFY requires null post-states and specific clarification items. Programs check reference formats, uniqueness, and message visibility. New requirements may appear only in post-request states.

Each judge call adds instructions, candidates, relation definitions, and the response contract below. Responses contain only schema\_version and relations and must cover exactly the requested candidates. The table lists request and response schema versions.

\begin{table}[htbp]
\caption{Semantic-judge request and response contracts.}
\label{tab:judge-contracts}
\centering\normalsize
\setlength{\tabcolsep}{3pt}
\renewcommand{\arraystretch}{1.08}
\begin{tabularx}{\linewidth}{@{}>{\raggedright\arraybackslash}p{0.23\linewidth}>{\raggedright\arraybackslash}p{0.46\linewidth}>{\raggedright\arraybackslash}X@{}}
\toprule
Judgment & \shortstack[c]{Request \(\rightarrow\)\\response\\version} & \shortstack[c]{Relation-item\\fields} \\
\midrule
RQ1 requirement relation & rq1-alignment-request-v1 \(\rightarrow\) rq1-alignment-response-v1 & prediction\_ref, gold\_ref, relation \\
RQ2/RQ3 requirement relation & requirement-alignment-request-v1 \(\rightarrow\) requirement-alignment-response-v1 & prediction\_ref, gold\_ref, relation \\
Attribute-field identity & state-field-alignment-request-v1 \(\rightarrow\) state-field-alignment-response-v1 & prediction\_field\_ref, gold\_field\_ref, relation \\
Text equivalence & state-semantic-request-v1 \(\rightarrow\) state-semantic-response-v1 & fact\_id, relation \\
Clarification blocker and question & rq3-clarification-semantic-request-v1 \(\rightarrow\) rq3-clarification-semantic-response-v1 & candidate\_id, issue\_relation, question\_validity \\
\bottomrule
\end{tabularx}
\end{table}

Requirement alignment covers every prediction--Gold pair using the labels in \ref{app:c1}. Attribute alignment considers unmatched fields within the same requirement after exact-path matching and judges variable meaning. Text requests cover only semantic comparisons specified by Gold comparators; programs handle numeric, enum, set, and missing-field comparisons. Clarification judging compares blockers within matched requirements and whether questions elicit the missing facts, retaining exact dimension and field checks. Text and blocker labels are EQUIVALENT, NOT\_EQUIVALENT, or UNCERTAIN; question labels are VALID, INVALID, or UNCERTAIN. The scoring program constructs these requests, checks versions, fields, scope, duplicates, and coverage, then calculates scores.

Experimental prompts and scoring code remain unchanged after the formal experiments. Diagnostic runs record prompt, instruction, and response-schema fingerprints and retain the four public inputs. Judge calls also retain requests, responses, and call records, with candidates, relation definitions, and response contracts in each request. Version checks use the saved records for the corresponding run.

\FloatBarrier
\section{Supplementary Results and Failure Case}
\label{app:e}

\FloatBarrier
\subsection{Complete Delivery Results}
\label{app:e1}

Each configuration and condition uses the same 37 targets, and means weight the six configurations equally. The C1/C2 results used in the Oracle comparisons are the same main-experiment results, with the same history conditions and input-output procedure. Pre-State and Post-State scores use a common reviewed validator version and combine harness-level scores with independent rescoring, including GPT-5.6 Sol results not copied back to the original summary.

Of the 222 Post-State configuration--target evaluations, 219 submit artifacts; 184 of these submissions pass build and regression checks but fail the target test, accounting for 84.02\% of submitted artifacts.

\begin{table}[htbp]
\centering\normalsize
\setlength{\tabcolsep}{3pt}
\renewcommand{\arraystretch}{1.08}
\caption{Complete-delivery counts and rates.}
\label{tab:complete-delivery-counts}
\begin{tabularx}{\linewidth}{@{}>{\raggedright\arraybackslash}Xrrrr@{}}
\toprule
Configuration & \shortstack[c]{C1 (Full\\History)} & \shortstack[c]{C2 (Relevant\\History)} & \shortstack[c]{Oracle\\Pre-State} & \shortstack[c]{Oracle\\Post-State} \\
\midrule
Codex / GPT-6 Astra & \shortstack[r]{4/37\\(10.81\%)} & \shortstack[r]{2/37\\(5.41\%)} & \shortstack[r]{2/37\\(5.41\%)} & \shortstack[r]{4/37\\(10.81\%)} \\
Codex / GPT-6 Sol & \shortstack[r]{2/37\\(5.41\%)} & \shortstack[r]{1/37\\(2.70\%)} & \shortstack[r]{1/37\\(2.70\%)} & \shortstack[r]{5/37\\(13.51\%)} \\
Codex / GPT-5.6 Sol & \shortstack[r]{1/37\\(2.70\%)} & \shortstack[r]{2/37\\(5.41\%)} & \shortstack[r]{4/37\\(10.81\%)} & \shortstack[r]{5/37\\(13.51\%)} \\
Codex / GPT-5.6 Terra & \shortstack[r]{2/37\\(5.41\%)} & \shortstack[r]{0/37\\(0.00\%)} & \shortstack[r]{1/37\\(2.70\%)} & \shortstack[r]{2/37\\(5.41\%)} \\
Claude Code / Opus 5.5 & \shortstack[r]{2/37\\(5.41\%)} & \shortstack[r]{1/37\\(2.70\%)} & \shortstack[r]{2/37\\(5.41\%)} & \shortstack[r]{2/37\\(5.41\%)} \\
Claude Code / Sonnet 5 & \shortstack[r]{0/37\\(0.00\%)} & \shortstack[r]{0/37\\(0.00\%)} & \shortstack[r]{0/37\\(0.00\%)} & \shortstack[r]{0/37\\(0.00\%)} \\
Mean & 4.95\% & 2.70\% & 4.50\% & 8.11\% \\
\bottomrule
\end{tabularx}
\end{table}

\begin{table}[htbp]
\centering\normalsize
\setlength{\tabcolsep}{3pt}
\renewcommand{\arraystretch}{1.08}
\caption{Oracle component counts and rates by configuration.}
\label{tab:oracle-component-counts}
\begin{tabularx}{\linewidth}{@{}>{\raggedright\arraybackslash}Xlrrrr@{}}
\toprule
Configuration & Condition & Build & Target & Regression & All three \\
\midrule
GPT-6 Astra & Pre & \shortstack[r]{18/37\\(48.65\%)} & \shortstack[r]{2/37\\(5.41\%)} & \shortstack[r]{15/37\\(40.54\%)} & \shortstack[r]{2/37\\(5.41\%)} \\
GPT-6 Astra & Post & \shortstack[r]{35/37\\(94.59\%)} & \shortstack[r]{5/37\\(13.51\%)} & \shortstack[r]{30/37\\(81.08\%)} & \shortstack[r]{4/37\\(10.81\%)} \\
GPT-6 Sol & Pre & \shortstack[r]{16/37\\(43.24\%)} & \shortstack[r]{2/37\\(5.41\%)} & \shortstack[r]{13/37\\(35.14\%)} & \shortstack[r]{1/37\\(2.70\%)} \\
GPT-6 Sol & Post & \shortstack[r]{35/37\\(94.59\%)} & \shortstack[r]{6/37\\(16.22\%)} & \shortstack[r]{30/37\\(81.08\%)} & \shortstack[r]{5/37\\(13.51\%)} \\
GPT-5.6 Sol & Pre & \shortstack[r]{23/37\\(62.16\%)} & \shortstack[r]{4/37\\(10.81\%)} & \shortstack[r]{21/37\\(56.76\%)} & \shortstack[r]{4/37\\(10.81\%)} \\
GPT-5.6 Sol & Post & \shortstack[r]{36/37\\(97.30\%)} & \shortstack[r]{5/37\\(13.51\%)} & \shortstack[r]{35/37\\(94.59\%)} & \shortstack[r]{5/37\\(13.51\%)} \\
GPT-5.6 Terra & Pre & \shortstack[r]{19/37\\(51.35\%)} & \shortstack[r]{1/37\\(2.70\%)} & \shortstack[r]{17/37\\(45.95\%)} & \shortstack[r]{1/37\\(2.70\%)} \\
GPT-5.6 Terra & Post & \shortstack[r]{34/37\\(91.89\%)} & \shortstack[r]{2/37\\(5.41\%)} & \shortstack[r]{33/37\\(89.19\%)} & \shortstack[r]{2/37\\(5.41\%)} \\
Opus 5.5 & Pre & \shortstack[r]{30/37\\(81.08\%)} & \shortstack[r]{2/37\\(5.41\%)} & \shortstack[r]{28/37\\(75.68\%)} & \shortstack[r]{2/37\\(5.41\%)} \\
Opus 5.5 & Post & \shortstack[r]{37/37\\(100.00\%)} & \shortstack[r]{2/37\\(5.41\%)} & \shortstack[r]{37/37\\(100.00\%)} & \shortstack[r]{2/37\\(5.41\%)} \\
Sonnet 5 & Pre & \shortstack[r]{28/37\\(75.68\%)} & \shortstack[r]{0/37\\(0.00\%)} & \shortstack[r]{26/37\\(70.27\%)} & \shortstack[r]{0/37\\(0.00\%)} \\
Sonnet 5 & Post & \shortstack[r]{37/37\\(100.00\%)} & \shortstack[r]{0/37\\(0.00\%)} & \shortstack[r]{37/37\\(100.00\%)} & \shortstack[r]{0/37\\(0.00\%)} \\
\bottomrule
\end{tabularx}
\end{table}

\textbf{Counts corresponding to the history conditions in Table~\ref{tab:oracle-delivery}.} The table expresses the pass rates in Table~\ref{tab:oracle-delivery} as passing-target counts out of 37 for each configuration, using the same C1/C2 conditions as the main experiment.

\begin{table}[htbp]
\caption{Component pass counts under C1 and C2 (37 targets per condition).}
\label{tab:history-component-counts}
\centering\normalsize
\setlength{\tabcolsep}{3pt}
\renewcommand{\arraystretch}{1.08}
\begin{tabularx}{\linewidth}{@{}>{\raggedright\arraybackslash}Xlrrrr@{}}
\toprule
Configuration & Condition & Build & Target & Regression & All three \\
\midrule
GPT-6 Astra & C1 & 21 & 6 & 17 & 4 \\
GPT-6 Astra & C2 & 16 & 3 & 14 & 2 \\
GPT-6 Sol & C1 & 18 & 3 & 17 & 2 \\
GPT-6 Sol & C2 & 12 & 1 & 9 & 1 \\
GPT-5.6 Sol & C1 & 22 & 1 & 19 & 1 \\
GPT-5.6 Sol & C2 & 17 & 2 & 14 & 2 \\
GPT-5.6 Terra & C1 & 20 & 2 & 18 & 2 \\
GPT-5.6 Terra & C2 & 14 & 0 & 10 & 0 \\
Opus 5.5 & C1 & 28 & 2 & 24 & 2 \\
Opus 5.5 & C2 & 27 & 1 & 26 & 1 \\
Sonnet 5 & C1 & 21 & 0 & 19 & 0 \\
Sonnet 5 & C2 & 19 & 0 & 17 & 0 \\
\bottomrule
\end{tabularx}
\end{table}

\textbf{Oracle process records and frozen submissions.} The table summarizes saved process and submission-freeze records for each run. Each configuration and condition uses 37 targets, totaling 222 target--configuration observations. Under Pre-State, 86 runs do not enter coding because the agent selects CLARIFY. Post-State has 219 process records and frozen submissions, while three runs have no frozen submission.

\begin{table}[htbp]
\caption{Oracle process and submission counts.}
\label{tab:oracle-submissions}
\centering\normalsize
\setlength{\tabcolsep}{3pt}
\renewcommand{\arraystretch}{1.08}
\begin{tabularx}{\linewidth}{@{}>{\raggedright\arraybackslash}Xrrrrrr@{}}
\toprule
Configuration & \shortstack[c]{Pre\\records} & \shortstack[c]{Pre\\submissions} & \shortstack[c]{Pre no\\submission} & \shortstack[c]{Post\\records} & \shortstack[c]{Post\\submissions} & \shortstack[c]{Post no\\submission} \\
\midrule
GPT-6 Astra & 19 & 19 & 18 & 37 & 37 & 0 \\
GPT-6 Sol & 17 & 17 & 20 & 37 & 37 & 0 \\
GPT-5.6 Sol & 23 & 23 & 14 & 37 & 37 & 0 \\
GPT-5.6 Terra & 19 & 19 & 18 & 34 & 34 & 3 \\
Opus 5.5 & 30 & 30 & 7 & 37 & 37 & 0 \\
Sonnet 5 & 28 & 28 & 9 & 37 & 37 & 0 \\
Total & 136 & 136 & 86 & 219 & 219 & 3 \\
\bottomrule
\end{tabularx}
\end{table}

\FloatBarrier
\subsection{Supplementary Requirement Diagnostic Results}
\label{app:e2}

Table~\ref{tab:diagnostic-selected} in the main text presents selected RQ1--RQ2 metrics under C1; the complete diagnostic results are reported below. RQ1 is evaluated only under Full History (C1). Requirement and evidence Precision, Recall, and F1 are computed per target and macro-averaged over 349 targets per configuration. RQ2 compares C1 and C2: Requirement Coverage is averaged over all 349 targets, whereas Attribute Reconstruction Scores exclude legitimate N/A cases and retain zero scores for agent failures. N denotes the number of targets included in each attribute-score mean.

\begin{table}[t]
\caption{Historical information identification and pre-request state reconstruction. (a) Requirement and evidence identification under Full History (C1). (b) Attribute reconstruction and requirement coverage; within each numeric cell, C1 is above C2. Coverage uses 349 tasks per configuration. Attribute scores exclude legitimate N/A cases and retain zero scores for agent failures. Attribute-score sample sizes are given in Table~\ref{tab:diagnostic-sample-sizes}.}
\label{tab:diagnostic-results-app}
\centering\normalsize
\setlength{\tabcolsep}{1.2pt}
\edef\diagrowheight{\the\dimexpr2\baselineskip\relax}
\edef\diagrowdepth{\the\dimexpr0.8\baselineskip\relax}
\newcommand{\diagrowstrut}{\rule[-\diagrowdepth]{0pt}{\diagrowheight}}
\newcommand{\diagpair}[2]{{\renewcommand{\arraystretch}{1}\begin{tabular}[c]{@{}r@{}}#1\\#2\end{tabular}}}
\begin{minipage}[t]{\linewidth}
\vspace{0pt}
\centering
(a) Historical information identification\par\smallskip
\begin{tabular*}{\linewidth}{@{\extracolsep{\fill}}lrrrrrr@{}}
\toprule
\multirow{2}{*}{Configuration} & \multicolumn{3}{c}{Requirement} & \multicolumn{3}{c}{Evidence} \\
\cmidrule(lr){2-4}\cmidrule(l){5-7}
 & P & R & F1 & P & R & F1 \\
\midrule
\diagrowstrut GPT-6 Astra & 0.2577 & 0.6821 & 0.3423 & 0.2254 & 0.9211 & 0.3166 \\
\diagrowstrut GPT-6 Sol & 0.3196 & 0.6704 & 0.3980 & 0.3395 & 0.8620 & 0.4207 \\
\diagrowstrut GPT-5.6 Sol & 0.3768 & 0.6302 & 0.4331 & 0.3162 & 0.8182 & 0.3939 \\
\diagrowstrut GPT-5.6 Terra & 0.4428 & 0.5232 & 0.4507 & 0.4254 & 0.6670 & 0.4416 \\
\diagrowstrut Claude Opus 5.5 & 0.3567 & 0.6232 & 0.4193 & 0.3189 & 0.8364 & 0.4048 \\
\diagrowstrut Claude Sonnet 5 & 0.4027 & 0.5835 & 0.4462 & 0.3060 & 0.7461 & 0.3766 \\
\bottomrule
\end{tabular*}
\end{minipage}\par\medskip
\begin{minipage}[t]{\linewidth}
\vspace{0pt}
\centering
(b) Pre-request state reconstruction\par\smallskip
\begin{tabular*}{\linewidth}{@{\extracolsep{\fill}}lrr@{}}
\toprule
\multirow{2}{*}{Configuration} & Attribute & Requirement \\
\cmidrule(lr){2-2}\cmidrule(l){3-3}
 & score & coverage \\
\midrule
GPT-6 Astra & \diagpair{0.2813}{0.3019} & \diagpair{0.6689}{0.6607} \\
GPT-6 Sol & \diagpair{0.2671}{0.3062} & \diagpair{0.6777}{0.6271} \\
GPT-5.6 Sol & \diagpair{0.2451}{0.3127} & \diagpair{0.6157}{0.5992} \\
GPT-5.6 Terra & \diagpair{0.2424}{0.2918} & \diagpair{0.4950}{0.5834} \\
Claude Opus 5.5 & \diagpair{0.2811}{0.3342} & \diagpair{0.5934}{0.5878} \\
Claude Sonnet 5 & \diagpair{0.2548}{0.2615} & \diagpair{0.5704}{0.5715} \\
\bottomrule
\end{tabular*}
\end{minipage}
\end{table}

\begin{table}[htbp]
\centering\normalsize
\setlength{\tabcolsep}{3pt}
\renewcommand{\arraystretch}{1.08}
\caption{Numbers of targets included in the Attribute Reconstruction Score means.}
\label{tab:diagnostic-sample-sizes}
\begin{tabular}{@{}lrr@{}}
\toprule
Configuration & N C1 & N C2 \\
\midrule
GPT-6 Astra & 265 & 266 \\
GPT-6 Sol & 277 & 252 \\
GPT-5.6 Sol & 251 & 249 \\
GPT-5.6 Terra & 221 & 251 \\
Claude Opus 5.5 & 253 & 246 \\
Claude Sonnet 5 & 238 & 241 \\
\bottomrule
\end{tabular}
\end{table}

RQ1 and RQ2 use the same historical requirement sets. We addressed relation-label disagreements from separate judge calls through automated detection, human adjudication, and metric recalculation, as described in \ref{app:c3}. A case recorded before adjudication is shown below.

\textbf{Decisions.} Values are C1/C2 pairs. With 312 Gold-ACT and 37 Gold-CLARIFY targets, always ACT yields accuracy 0.8940 and balanced accuracy 0.5.

\begin{table}[htbp]
\caption{Action-selection performance under C1/C2. Paired entries show C1 above C2.}
\label{tab:action-selection}
\centering\normalsize
\setlength{\tabcolsep}{3pt}
\renewcommand{\arraystretch}{1.08}
\begin{tabular}{@{}lrrrr@{}}
\toprule
Configuration & \shortstack[c]{Decision\\Accuracy} & \shortstack[c]{Balanced\\Accuracy} & ACT Recall & CLARIFY Recall \\
\midrule
GPT-6 Astra & \shortstack[r]{0.5931\\0.4670} & \shortstack[r]{0.6771\\0.6066} & \shortstack[r]{0.5705\\0.4295} & \shortstack[r]{0.7838\\0.7838} \\
GPT-6 Sol & \shortstack[r]{0.5043\\0.3954} & \shortstack[r]{0.6156\\0.5666} & \shortstack[r]{0.4744\\0.3494} & \shortstack[r]{0.7568\\0.7838} \\
GPT-5.6 Sol & \shortstack[r]{0.5358\\0.4327} & \shortstack[r]{0.6094\\0.5874} & \shortstack[r]{0.5160\\0.3910} & \shortstack[r]{0.7027\\0.7838} \\
GPT-5.6 Terra & \shortstack[r]{0.5415\\0.4413} & \shortstack[r]{0.6483\\0.6041} & \shortstack[r]{0.5128\\0.3974} & \shortstack[r]{0.7838\\0.8108} \\
Claude Opus 5.5 & \shortstack[r]{0.7880\\0.7335} & \shortstack[r]{0.6670\\0.6961} & \shortstack[r]{0.8205\\0.7436} & \shortstack[r]{0.5135\\0.6486} \\
Claude Sonnet 5 & \shortstack[r]{0.6361\\0.5673} & \shortstack[r]{0.6774\\0.6270} & \shortstack[r]{0.6250\\0.5513} & \shortstack[r]{0.7297\\0.7027} \\
\bottomrule
\end{tabular}
\end{table}

\textbf{Branch content.} Post-State uses 312 Gold-ACT targets; blocker, question, and strict-clarification metrics use 37 Gold-CLARIFY targets, with wrong branches scoring zero. Exact Post-State Accuracy is 0/312 for all six configurations under C1/C2.

\begin{table}[htbp]
\caption{Post-request state and clarification scores under C1/C2. Paired entries show C1 above C2.}
\label{tab:branch-content}
\centering\normalsize
\setlength{\tabcolsep}{3pt}
\renewcommand{\arraystretch}{1.08}
\begin{tabular}{@{}lrrrr@{}}
\toprule
Configuration & Post-State & \shortstack[c]{Blocking-Issue\\F1} & \shortstack[c]{Question\\Validity} & \shortstack[c]{Strict\\Clarification\\Success Rate} \\
\midrule
GPT-6 Astra & \shortstack[r]{0.0741\\0.0744} & \shortstack[r]{0.0874\\0.0667} & \shortstack[r]{0.1622\\0.1351} & \shortstack[r]{0.0000\\0.0000} \\
GPT-6 Sol & \shortstack[r]{0.0535\\0.0600} & \shortstack[r]{0.1051\\0.0667} & \shortstack[r]{0.2162\\0.1351} & \shortstack[r]{0.0000\\0.0000} \\
GPT-5.6 Sol & \shortstack[r]{0.0538\\0.0569} & \shortstack[r]{0.0739\\0.0802} & \shortstack[r]{0.1622\\0.1622} & \shortstack[r]{0.0000\\0.0000} \\
GPT-5.6 Terra & \shortstack[r]{0.0653\\0.0870} & \shortstack[r]{0.0676\\0.0901} & \shortstack[r]{0.1351\\0.1351} & \shortstack[r]{0.0000\\0.0270} \\
Claude Opus 5.5 & \shortstack[r]{0.1670\\0.1710} & \shortstack[r]{0.0423\\0.1523} & \shortstack[r]{0.0811\\0.2703} & \shortstack[r]{0.0000\\0.0000} \\
Claude Sonnet 5 & \shortstack[r]{0.1041\\0.1064} & \shortstack[r]{0.0694\\0.0874} & \shortstack[r]{0.1081\\0.1351} & \shortstack[r]{0.0270\\0.0270} \\
\bottomrule
\end{tabular}
\end{table}

\textbf{State quality given correct ACT.} The following reports conditional means and counts for Gold-ACT targets predicted ACT.

\begin{table}[htbp]
\caption{Post-State Scores conditional on correct ACT decisions.}
\label{tab:conditional-post-state}
\centering\normalsize
\setlength{\tabcolsep}{3pt}
\renewcommand{\arraystretch}{1.08}
\begin{tabular}{@{}lrrrr@{}}
\toprule
Configuration & \shortstack[c]{Correct ACT\\count\\C1} & \shortstack[c]{Conditional\\Post-State\\C1} & \shortstack[c]{Correct ACT\\count\\C2} & \shortstack[c]{Conditional\\Post-State\\C2} \\
\midrule
GPT-6 Astra & 178 & 0.1298 & 134 & 0.1733 \\
GPT-6 Sol & 148 & 0.1129 & 109 & 0.1719 \\
GPT-5.6 Sol & 161 & 0.1043 & 122 & 0.1456 \\
GPT-5.6 Terra & 160 & 0.1273 & 124 & 0.2189 \\
Claude Opus 5.5 & 256 & 0.2036 & 232 & 0.2300 \\
Claude Sonnet 5 & 195 & 0.1665 & 172 & 0.1929 \\
\bottomrule
\end{tabular}
\end{table}

\textbf{Set and content decomposition.} A requirement set fully matches when predicted, Gold, and matched counts satisfy $p=g=m>0$; Post-State then equals the mean matched-state score. To control target changes caused by C1/C2 action selection, the paired comparison uses the intersection of correct-ACT targets.

\begin{table}[htbp]
\caption{Requirement-set matching and shared-target Post-State Scores under C1/C2. Paired entries show C1 above C2.}
\label{tab:matched-state-diagnostics}
\centering\normalsize
\setlength{\tabcolsep}{3pt}
\renewcommand{\arraystretch}{1.08}
\begin{tabular}{@{}lrrrr@{}}
\toprule
Configuration & \shortstack[c]{Matched-set\\count C1 / C2} & \shortstack[c]{Matched-set\\Post C1 / C2} & \shortstack[c]{Shared ACT\\count} & \shortstack[c]{Shared-target\\Post C1 / C2} \\
\midrule
GPT-6 Astra & \shortstack[r]{50\\41} & \shortstack[r]{0.2882\\0.3715} & 127 & \shortstack[r]{0.1384\\0.1710} \\
GPT-6 Sol & \shortstack[r]{36\\38} & \shortstack[r]{0.2422\\0.3492} & 99 & \shortstack[r]{0.1317\\0.1591} \\
GPT-5.6 Sol & \shortstack[r]{39\\40} & \shortstack[r]{0.2299\\0.3215} & 107 & \shortstack[r]{0.1051\\0.1528} \\
GPT-5.6 Terra & \shortstack[r]{46\\41} & \shortstack[r]{0.2899\\0.4563} & 100 & \shortstack[r]{0.1471\\0.2358} \\
Claude Opus 5.5 & \shortstack[r]{79\\68} & \shortstack[r]{0.3656\\0.4468} & 221 & \shortstack[r]{0.2082\\0.2320} \\
Claude Sonnet 5 & \shortstack[r]{54\\57} & \shortstack[r]{0.3320\\0.3604} & 138 & \shortstack[r]{0.1709\\0.2029} \\
\bottomrule
\end{tabular}
\end{table}

Within the correct-ACT subsets, the fully matched-set proportions averaged across configurations are 27.32\%/32.29\%, with subset Post-State means 0.2913/0.3843 and 304/285 target\(\times\)configuration observations. The shared correct-ACT subset contains 792 observations, with the configuration mean rising from 0.1503 to 0.1923. Statistics are computed within configurations before equal-weight averaging; pooled observations are not independent targets.

\textbf{Cross-RQ label disagreement before human adjudication.} For 42204309\_T007, two judge calls assigned different labels to the same prediction--Gold requirement pair. The table retains the labels and scores before adjudication to illustrate the detected conflict. Recall and Coverage are target-level metrics.

\begin{table}[htbp]
\caption{Cross-diagnostic label disagreement before human adjudication.}
\label{tab:judge-disagreement}
\centering\normalsize
\setlength{\tabcolsep}{3pt}
\renewcommand{\arraystretch}{1.08}
\begin{tabularx}{\linewidth}{@{}>{\raggedright\arraybackslash}p{0.13\linewidth}>{\raggedright\arraybackslash}p{0.16\linewidth}>{\raggedright\arraybackslash}X>{\raggedright\arraybackslash}p{0.23\linewidth}>{\raggedright\arraybackslash}p{0.22\linewidth}@{}}
\toprule
Evaluation & \shortstack[c]{Prediction\\reference} & \shortstack[c]{Gold\\requirement} & Relation & Target metric \\
\midrule
RQ1 & \path{reward_payout} & Prize Claim Flow & MERGED\_ATOMS & Requirement Recall = 0 \\
RQ2 & \path{reward_payout} & Prize Claim Flow & SAME\_ATOM & Requirement Coverage = 0.25 \\
\bottomrule
\end{tabularx}
\end{table}

After the program detected this disagreement, human reviewers resolved it using the requirement-atom definition and source evidence. The program then updated matching and recalculated the affected metrics.

\textbf{Decomposition of Exact Post-State Accuracy.} Using only saved frozen scores, we assign each Gold-ACT target to the first unmet condition in the order of action, requirement set, state fields, and removed-attribute list. Categories are disjoint and sum to 312 under each history condition. The first category includes wrong action decisions and incomplete responses. The removal-list category applies only after all earlier conditions pass, so zero there does not imply correct lists on all other targets.

\begin{table}[htbp]
\caption{Decomposition of exact post-request state failures under C1/C2.}
\label{tab:exact-state-failures}
\centering\normalsize
\setlength{\tabcolsep}{3pt}
\renewcommand{\arraystretch}{1.08}
\begin{tabular}{@{}lrrrrr@{}}
\toprule
Configuration & \shortstack[c]{No correct\\ACT} & \shortstack[c]{Set\\mismatch} & \shortstack[c]{State not\\fully exact} & \shortstack[c]{Removal list\\only} & Exact success \\
\midrule
GPT-6 Astra & 134 / 178 & 128 / 93 & 50 / 41 & 0 / 0 & 0 / 0 \\
GPT-6 Sol & 164 / 203 & 112 / 71 & 36 / 38 & 0 / 0 & 0 / 0 \\
GPT-5.6 Sol & 151 / 190 & 122 / 82 & 39 / 40 & 0 / 0 & 0 / 0 \\
GPT-5.6 Terra & 152 / 188 & 114 / 83 & 46 / 41 & 0 / 0 & 0 / 0 \\
Claude Opus 5.5 & 56 / 80 & 177 / 164 & 79 / 68 & 0 / 0 & 0 / 0 \\
Claude Sonnet 5 & 117 / 140 & 141 / 115 & 54 / 57 & 0 / 0 & 0 / 0 \\
\bottomrule
\end{tabular}
\end{table}

Each cell reports C1 / C2 target counts. All 589 saved scores with fully matched requirement sets fail at least one strict state-correctness condition; no target fails only the final removal-list condition. This explains the zero strict accuracy, while the existing continuous scores retain partial credit. The audit generated no new responses and changed neither semantic labels nor existing scores.

\FloatBarrier
\subsection{PCB Failure Case}
\label{app:e3}

\textbf{PCB case: Relevant requirements are found, but the target state remains incomplete.} Under C1 on \path{43214420_T004}, the client asks to move the battery connector to the back of the board, on the side opposite the USB-C connector. Opus 5.5 identifies both relevant historical requirements, achieving requirement Precision and Recall of 1, evidence Recall of 1, evidence Precision of 0.625, and RQ2 Requirement Coverage of 1. Its pre-request state already includes layout constraints such as 8 mounting holes and the M2.5 specification. However, its post-request response retains only the battery connector and omits the entire mechanical-layout requirement, yielding a Post-State Score of 0.3259. The final delivery passes the build and regression checks but fails the task-specific test. This record shows the co-occurrence of historical coverage, target-state omission, and delivery failure; the case alone does not establish causality. The continuous score is also affected by comparisons of other battery-connector fields, whereas omission of the entire mechanical-layout requirement can be checked directly from the prediction structure.

This case comes from the benchmark's linked two-phase evaluation and illustrates omission of a valid constraint within one response, together with the corresponding delivery checks.

\textbf{Oracle artifact-implementation case.} GPT-5.6 Sol's 43214420\_T004 comes from a separate Oracle Post-State run, not coding continued from the Opus diagnostic response above. The structured artifact model (\path{artifact_model}) and design summary update battery-connector placement to the back opposite USB-C, but the delivered PCB file is byte-identical to the initial archive. Build and regression pass; the target test reports that the connector-edge normals are not opposite.

\textbf{State and artifact comparison.} The tables correspond to the two independent runs described above. The first examines one requirement response; the second examines the Oracle deliverable.

\begin{table}[htbp]
\caption{Requirement-state omission in the Opus diagnostic case.}
\label{tab:opus-state-case}
\centering\normalsize
\setlength{\tabcolsep}{3pt}
\renewcommand{\arraystretch}{1.08}
\begin{tabularx}{\linewidth}{@{}>{\raggedright\arraybackslash}X>{\raggedright\arraybackslash}X@{}}
\toprule
\shortstack[c]{Opus\\diagnostic\\record} & \shortstack[c]{Saved field\\and result} \\
\midrule
Pre-request mechanical requirement & requirement\_ref = req\_power\_board\_outline; mounting\_holes = "8 at the corners"; mounting\_hole\_size = "M2.5" \\
Post-request references & only req\_battery\_connector\_placement; no req\_power\_board\_outline \\
Identification and coverage & Requirement Precision = 1; Recall = 1; RQ2 Coverage = 1 \\
Post-request state & Post-State Score = 0.325926; mechanical requirement absent \\
Checks & Build PASS; Target FAIL; Regression PASS \\
Target error & connector edge normals are not opposite (dot=0.000000) \\
\bottomrule
\end{tabularx}
\end{table}

\begin{table}[htbp]
\caption{Artifact changes and delivery checks in the Oracle PCB case.}
\label{tab:oracle-pcb-case}
\centering\normalsize
\setlength{\tabcolsep}{3pt}
\renewcommand{\arraystretch}{1.08}
\begin{tabularx}{\linewidth}{@{}>{\raggedright\arraybackslash}X>{\raggedright\arraybackslash}X@{}}
\toprule
\shortstack[c]{GPT-5.6 Sol\\Oracle\\artifact} & \shortstack[c]{File\\comparison and\\checks} \\
\midrule
Changed files & Design summary, artifact model, current-state representation, and test files \\
Delivered PCB & Byte-identical to initial archive \\
Checks & Build PASS; Target FAIL; Regression PASS \\
Target error & connector edge normals are not opposite (dot=1.000000) \\
\bottomrule
\end{tabularx}
\end{table}

\FloatBarrier
The omission in the first run can be checked directly from its output items. Comparing the initial and final archives in the second run distinguishes changes to structured descriptions from changes to the delivered PCB artifact.

\FloatBarrier
\end{document}